\documentclass[a4paper,11pt]{article}
\pdfoutput=1 

\usepackage{jheppub} 

\usepackage[T1]{fontenc} 

\usepackage{xparse}

\usepackage{appendix}
\usepackage{graphicx}
\graphicspath{ {./Figures/} }

\usepackage{dcolumn}
\usepackage{bm}
\usepackage{hyperref}
\usepackage{xcolor} 

\usepackage{comment}

\newcommand{\Tr}{\text{Tr}}
\newcommand{\tr}{\text{tr}}
\usepackage{slashed}
\renewcommand{\vec}[1]{\mathbf{#1}}

\usepackage{bm}
\newcommand{\vect}[1]{\boldsymbol{\mathbf{#1}}}

\usepackage{revsymb}

\usepackage{bbold}

\title{On the path integral approach to quantum anomalies in interacting models}

    \author[a]{Alireza Parhizkar,}
	\author[b]{Colin Rylands}
    \author[a,c]{and Victor Galitski}

\affiliation[a]{Joint Quantum Institute, University of Maryland, \\ College Park, MD 20742, USA}
\affiliation[b]{SISSA and INFN, \\ via Bonomea 265, 34136 Trieste ITALY}
\affiliation[c]{Center for Computational Quantum Physics, The Flatiron Institute, New York, NY 10010, United States}

\emailAdd{alpa@umd.edu}
\emailAdd{crylands@sissa.it}
\emailAdd{galitski@umd.edu}

\date{
    \today
}

\abstract{
The prediction and subsequent discovery of topological semimetal phases of matter in solid state systems has instigated a surge of activity investigating the exotic properties of these unusual materials. Amongst these are transport signatures which can be attributed to the chiral anomaly; the breaking of classical chiral symmetry in a quantum theory. 
 This remarkable quantum phenomenon,  first discovered in the context of particle physics has now found new life in condensed matter physics, connecting topological quantum matter and band theory with effective field theoretic models. 
In this paper we investigate the interplay between interactions and the chiral anomaly in field theories inspired by semimetals using Fujikawa’s path integral method. Starting from models in one spatial dimension we discuss how the presence of interactions can affect the consequences of the chiral anomaly leading to renormalization of excitations and their transport properties. This is then generalised to the three dimensional case where we show that the anomalous response of the system, namely the chiral magnetic and quantum hall effects, are modified by the presence of interactions. These properties are investigated further through the identification of anomalous  modes which exist within interacting Weyl semimetals. These massive excitations are nonperturbative in nature and are a direct consequence of the chiral anomaly. 
The effects of interactions on mixed axial-gravitational anomalies are then investigated and the conditions required for interactions effects to be observed are discussed.}

\begin{document}

\maketitle
\flushbottom


\section{Introduction}

The state of a physical system is defined through parameters that fully determine its configuration. Each distinct realization of these set of parameters corresponds to a distinct configuration. For example, revaluation of generalized coordinates $\vec{x}=(x,y,z)$ of a point particle takes it to different positions in the three dimensional space. A transformation $\vec{x} \rightarrow \mathcal{T}[\vec{x}]$ then, yields a different physical situation and it will be a symmetry only if all such situations are governed by the same physical laws. In other words, if $\mathcal{T}$ is a symmetry transformation, the physical description (or physics) of the system, will be blind to the difference between $\vec{x}$ and $\mathcal{T}[\vec{x}]$. For example, given a $\mathcal{T}$ that preserves the action functional, if the physical path $\vec{x_\text{ph}}$ minimizes the action then also will $\mathcal{T}[\vec{x_\text{ph}}]$.

There are two prominent types of physical descriptions, classical and quantum. Since the quantum description is canonically derived from the classical, one expects two situations that are classically indistinguishable, to remain indistinguishable in the quantum description as well. Therefore, it sounds peculiar that in some cases, a symmetry of the classical theory is not a symmetry of the corresponding quantum theory. It means that quantum corrections know something about the system that the classical description is completely ignorant about. This phenomena is known as a quantum anomaly; when the classical description of two situations are the same but their quantum descriptions differ. 

For every symmetry there is a corresponding conserved Noether current. Thus, anomalies which destroy a classical symmetry consequently ruin a classical conservation law, giving rise to a source term which is usually called the anomalous term. This non-vanishing source term is responsible for the decay of the neutral pion---the phenomenon which would have been suppressed if not for quantum anomalies and thus led to their discovery~\cite{MesonDecay,DecaySchwinger,Adler,BellJackiw,BardeenWI}.

Anomalies are hence purely quantum mechanical and therefore become of even greater interest when one finds out that there are macroscopic phenomena based on these beings which dwell deep in the quantum realm. One example of such macroscopic phenomena are anomalous transport signatures in condensed matter systems.

Quantum anomalies are among many that have originated in high energy physics but have gradually found their way into condensed matter physics. In this paper we will be concerned with the chiral anomaly and its corresponding anomalous term which supplies the non-conservation of the chiral current. In condensed matter systems such as crystals, chiral symmetry is an emergent property, since the periodic nature of the crystal results in a periodic band structure which allows for exactly as many left-handed chiral modes as there are right-handed ones. In other words, if a band crosses the Fermi surface at one point it will cross the Fermi surface back at least at one other point due to the periodicity of the band structure~\cite{NN1,NN2,NNnogo,NNFriedan}. As a consequence, exciting the system will generate left-moving particles around one point while it produces right-moving particles around the other. So a definite chirality cannot be attributed to a specific band. Even though chiral points are connected through the deep lattice structure, they can be thought of as distinct nodes in the low energy description of the material which is blind to the lattice structure. In this sense the chiral anomaly has a prosaic explanation in condensed matter systems, namely, the pumping of charge through the bottom of the band from one node to another. Amazingly, this simple picture relates key concepts such as the quantized Hall conductance (e.g. via Laughlin's argument~\cite{Laughlin}) and existence of topological metals (e.g. the Weyl semimetal~\cite{WanTurnerVishwanathSavrasov,BurkovBalents, YangLuRab, XuWengWangDaiFang, HalaszBalents, Aji,WengFangZhongBernevigDai, Lv1, Lv2, Xu, Huang}) to chiral anomalies.

Topology is the cornerstone of chiral anomaly. We will see that the chiral anomaly is connected to zero-modes of the fermionic theory on one hand and the winding number of the gauge field on the other, when we review the non-perturbative formulation of anomalies named after Fujikawa~\cite{Fujikawa, FujikawaErrata,Fujikawa2004Book} in section \ref{sec:Review}. Within perturbative QED the chiral anomaly arises \textit{only} from the triangle diagrams when one needs to carefully regularize the difference between linearly divergence integrals~\cite{AdlerBardeen}. Higher order loop diagrams will cancel their own contribution to chiral symmetry breaking and render the chiral anomaly subject to non-renormalization theorems; the higher order terms will leave the form of the anomalous term unmodified and are accounted for by replacing the bare fields and parameters with their renormalized values.

While in theories of elementary particle physics, one is restricted by normalizability conditions and symmetries such as Lorentz invariance and charge conservation, this is not the case of condensed matter systems where we often seek effective low energy description of complex systems. Condensed matter physics is therefore home to a diverse array of particles and interactions, and thus low energy description of a condensed matter system is apt to carrying interaction terms that could be alien to fundamental physics. In this article we investigate the interplay of the chiral anomaly with these features that arise in condensed matter systems. In particular, building on a previous work \cite{CAICMS} we study the effects of interactions on anomalous chiral symmetry breaking in low energy descriptions of Dirac and Weyl materials.

After the review of the basics presented in section \ref{sec:Review}, we employ the path-integral formulation in section \ref{sec:2D} to investigate in detail the aspects of introducing interactions on chiral anomaly in a $(1+1)$-dimensional system e.g. a Luttinger liquid. We then proceed to do the same in the next section \ref{sec:4D} for $(3+1)$-dimensions where a much richer behaviour will emerge. As will we see, due to interactions there are modifications to the chiral symmetry breaking in both systems, and how they related to each other is the subject of section \ref{sec:DimRed}. As mentioned before there are macroscopic phenomena attributed to chiral anomaly such as chiral magnetic effect and anomalous Hall response. Since the anomalous term will be different in a theory that contains the additional interaction, these phenomena also will be modified by interactions. In particular we will see that even though Hall conductivity will remain the same in the equilibrium and homogeneous limit, it will have a different finite frequency behavior in presence of interactions. These are investigated in section \ref{sec:Measurement} where we have represented them as a way to measure the effects of interactions. Curiously, the effects of interaction exceed the mere modification of anomalous transport phenomena; in section \ref{sec:DynamicAnomaly} we will see that duo to the interplay of interactions and Weyl node separation, anomalous current will have a dynamic of its own even in the absence of electromagnetic gauge field. These ``anomalous modes'' can further enforce dynamics on the gauge field; in particular they give rise to an axionic electrodynamics. Section \ref{sec:Gravity} carries on to thermal phenomena and investigates how interactions influence the gravitational anomaly and transport in presence of non-trivial geometry. Finally, although we are mainly considering the consequences of a general local current-current interactions in this paper, in section \ref{sec:Beyond} we go beyond this type and consider one other kind which also can encapsulate the effect of local spin-spin interactions on chiral anomaly.

Overall, a reader who is \textit{only} interested in the effect of interactions on anomaly in $(3+1)$ dimensions and is familiar with the subject of chiral anomaly and Fujikawa's method for deriving it, may start from section \ref{sec:4D} and go back to previous sections upon questions or difficulties.

\subsection*{Notations}

Throughout the paper we use the natural units where the reduced Planck constant $\hbar$ and the speed of light $c$ are set to one. We also use Einstein's summation convention and sometimes represent the four-vector of current as $j^\mu \equiv (\rho,j^x,j^y,j^z)$ in Minkowski coordinates $x^\mu \equiv (t,x,y,z)$. The Minkowskian metric is denoted by $\eta_{\mu\nu} = \text{diag} (1,-1,-1,-1)$, the Kronecker delta by $\delta^\mu_\nu$, the d'Alembertian operator by  $\Box \equiv \partial_t^2 - \vect\nabla^2$ and Dirac's slash notation, $\gamma^\mu \mathcal{V}_\mu = \slashed{\mathcal{V}}$, is employed.

Gamma matrices, $\gamma^\mu$, are the matrix representations of the Clifford algebra: $\{\gamma^\mu,\gamma^\nu\}=2\eta^{\mu\nu}$. We also define $\gamma_5$ as the matrix that satisfies a natural extension $\{\gamma^\mu,\gamma_5\}=0$ and $\gamma_5^2=1$. In a two dimensional spacetime we can choose the representation as $\gamma^0 = \sigma^x$ and $\gamma^1=i\sigma^y$, yielding $\gamma_5=\sigma^z$, with $\sigma^{x,y,z}$ being the Pauli matrices. In four dimensions we can write them as,
\begin{equation}
    \gamma^\mu = \left( \begin{array}{cc}
         0 & \sigma^\mu  \\
        \bar\sigma^\mu & 0
    \end{array} \right) \, ,
    \ \quad \ 
    \gamma_5  = \left( \begin{array}{cc}
         -\mathbb{1} & 0  \\
        0 & \mathbb{1}
    \end{array} \right) \, ,
\end{equation}
with $\sigma^\mu \equiv (\mathbb{1},\vect{\sigma})$ and $\bar\sigma^\mu \equiv (\mathbb{1},-\vect{\sigma})$. The basis of the above particular representation is called the Weyl basis. A Wick rotation, $t\rightarrow -it$, changes the Minkowskian metric, $\eta_{\mu\nu} \rightarrow \text{diag} (-1,-1,-1,-1)$, and hence changes the algebra, which leads to a redefinition of the zeroth gamma matrix by $\gamma^0 \rightarrow -i\gamma^0$.

Moreover, to avoid clutter and better compatibility with high energy physics, we have treated the Fermi velocity as the speed of light unless when it was needed to restate it distinctly.

\section{Fujikawa's Method} \label{sec:Review}

Here we briefly review our main technical tool, Fujikawa's path-integral approach to calculating quantum anomalies. The reader who already has a fresh knowledge of the subject can safely pass through to the next section.

\subsection{Non-Trivial Jacobian}

A symmetry transformation leaves the equations of motion unchanged, or in the language of action principle, it preserves the action functional (up to a boundary term). Let us assume action functional $S[\Phi]$ describes our classical system and it has as symmetry transformation $\mathcal{T}$ that preserves it. All the degrees of freedom are represented by $\Phi$. Let us further formulate our quantum theory by the path-integral approach. All there is to know about the quantum system are given by path-integrals of the form $I=\int \mathcal{D}\Phi e^{iS[\Phi]}$. An anomaly, as we have introduced it, is an instance when a classical symmetry fails to be also a quantum one. But $S$ in invariant under $\mathcal{T}$, so $I$ also would be if it was not for the path-integral measure $\mathcal{D}\Phi$. The only thing that can ruin the symmetry at the quantum level, when described by the path-integral language, is evidently the measure. Thus for the quantum anomaly to appear $\mathcal{D}\Phi$ must transform under $\mathcal{T}$ with a non-trivial Jacobian of transformation. This remarkable transparency, where the quantum anomaly is \textit{anticipated} at the outset, is a feature of the path-integral approach in comparison with the perturbative approach where the quantum anomaly is \textit{discovered} when higher order corrections are being calculated. This Lagrangian formalism of quantum anomalies is named after Fujikawa~\cite{Fujikawa, FujikawaErrata,Fujikawa2004Book}. 

In this paper we are concerned with chiral anomalies that happen in fermionic systems. For concreteness consider the following action functional in (1+1)-dimensional space-time having sufficient dimensions to capture the essence of the quantum anomalies,
\begin{equation}
	S[\bar{\psi},\psi,A_\mu] = \int d^2 x \left[ \bar{\psi} i\gamma^\mu \left( \partial_\mu - i e A_\mu \right) \psi \right] \,  ,
	\label{SimpleAction}
\end{equation}
along with the corresponding path-integral,
\begin{equation}
	I = \int \mathcal{D}\bar{\psi}\mathcal{D}\psi e^{iS[\bar{\psi},\psi,A_\mu]} \, ,
\end{equation}
where the integration is only over fermionic degrees of freedom $\psi$ and $\bar\psi \equiv \psi^\dagger \gamma^0$ (which in the current case are Weyl spinors) hence treating the gauge field $A^\mu$ as an external electromagnetic four-potential. Note also, that the path-integral treats $\psi$ and $\bar\psi$ as independent integral variables.

The action \eqref{SimpleAction} is symmetric under two $U(1)$ global transformations. A simple phase transformation,
\begin{equation}
	\psi \longrightarrow e^{i\alpha} \psi \, , \ \ \bar{\psi} \longrightarrow \bar{\psi} e^{-i\alpha} \, ,
	\label{PhaseTransG}
\end{equation}
and a chiral transformation,
\begin{equation}
	\psi \longrightarrow e^{i\alpha\gamma_5} \psi \, , \ \ \bar{\psi} \longrightarrow \bar{\psi} e^{i\alpha\gamma_5} \, .
	\label{ChiralTransG}
\end{equation}
We are then curious to see how the measure $\mathcal{D}\bar{\psi}\mathcal{D}\psi$ behaves under the above transformations.

To begin our investigation we use the usual method of Euclideanization, which has the reward of making the Dirac operator $\slashed{D} \equiv \gamma^\mu (\partial_\mu - ie A_\mu)$ hermitian and helps significantly in the calculation of path-integrals. Hermitian operators can form complete orthonormal bases. Therefore, if now we decide to expand the fermionic fields, there is a natural way of doing so. Namely, we can expand $\bar{\psi}$ and $\psi$ in orthonormal modes of the hermitian Dirac operator $\slashed{D}\phi_n = l_n \phi_n$, with $l_n$s being eigenvalues of $\slashed{D}$ and $\phi_n$s the corresponding eigenfunctions.
\begin{equation}
	\bar{\psi} = \sum_n \bar{b}_n \phi^\dagger_n(x) \, , \ \ \psi=\sum_n a_n\phi_n(x) \, .
	\label{SpinorExpansion}
\end{equation}
Both $a_n$ and $\bar{b}_n$ are Grassmann numbers so that they form Grassmann field spinors when multiplied by two component fields $\phi_n$ and $\phi^\dagger_n$.

We could have expanded the fermionic fields in the basis of any hermitian operator. But this natural choice has the crucial property of formally diagonalizing the action. By this choice it becomes clear that the fermionic part of the path-integral is given by the products of all eigenvalues of the Dirac operator which renders it exactly integrable.
\begin{align}
	I &=  \int \mathcal{D}\bar{\psi}\mathcal{D}\psi e^{S[\bar{\psi},\psi,A_\mu]} = \int \prod_n \left[ d\bar{b}_n da_n e^{l_n \bar{b_n}a_n} \right] = \prod_n l_n = \det (\slashed{D})
\end{align}
In above and future path-integral equalities we use the equal sign for all path-integrals that are proportional to each other by a constant coefficient. We should also note that integration over Grassmann numbers is defined by left derivative.

Other than orthonormality and completeness,
\begin{align}
	&\int d^2x \phi^\dagger_m(x) \phi_n(x) = \delta_{mn} 	\label{Orthonormality} \, , \\
	&\sum_n \phi^\dagger_n(x)\phi_n(y) = \delta(x-y) \, ,	\label{Completeness}
\end{align}
eigenfunctions of the Dirac operator have another property which is of interest here. Since $\slashed{D}$ anti-commutes with $\gamma_5$, multiplying an eigenfunction with $\gamma_5$ produces another eigenfunction of the Dirac operator with an eigenvalue negative of the the original one:
\begin{equation}
	\slashed{D}(\gamma_5\phi_n) = -\gamma_5\slashed{D}\phi_n = -l_n (\gamma_5\phi_n) \, .
	\label{MirroredMode}
\end{equation}
Therefore, if $l_n \neq 0$ then $\phi_n$ and $\gamma_5 \phi_n$ are orthogonal to each other. When $l_n=0$ the corresponding eigenfunction is called a zero mode. In the subset of zero modes $\slashed{D}$ and $\gamma_5$ can be simultaneously diagonalized, since $[\slashed{D},\gamma_5]\phi_n$ vanishes there.

By an infinitesimal chiral rotation \eqref{ChiralTransG} the expansion coefficients $a_n$ and $\bar{b}_n$ go under a transformation.
\begin{align}
	 \sum_n a_n\phi_n &\rightarrow \sum_n e^{i\alpha\gamma_5}  a_n\phi_n  \approx \sum_n (1 + i\alpha\gamma_5) a_n \phi_n \Rightarrow \nonumber \\
     a_m &\rightarrow a_m + i\sum_n a_n \int d^2x  \phi^\dagger_m(x) \alpha \gamma_5 \phi_n(x) \, ,
\end{align}
where we have used \eqref{Orthonormality} to go from first to second line and we have kept $\alpha$ inside the integral since in general it can depend on position. Therefore, $\{a_n\}$ transforms as $\{a_n\}' = M \{a_n\}$ with the matrix of transformation given by $M_{mn} = \left[ \delta_{mn} + i\int d^2x  \phi^\dagger_m(x) \alpha \gamma_5 \phi_n(x) \right]$. The exact same goes for $\bar{b}_n$ because a chiral rotation \eqref{ChiralTransG} transforms $\bar{\psi}$ the same way it transforms $\psi$, unlike a phase rotation \eqref{PhaseTransG}.
\begin{equation}
	\bar{b}_m \rightarrow \sum_n \left[ \delta_{mn} + i\int d^2x  \phi^\dagger_m(x) \alpha \gamma_5 \phi_n(x) \right] \bar{b}_n \, .
\end{equation}
Phase transformation is obtained from chiral transformation if we substitute $\gamma_5$ by $1$ everywhere, and $i$ by $-i$ only for variables that have a bar sign i.e. $\bar{\psi}$ and $\bar{b}_n$.

Since $a_n$ and $\bar{b}_n$ are Grassmann numbers and their integrals are given by left derivatives, the Jacobian of their transformation is the inverse of Jacobian for standard c-number transformations and is given by,
\begin{align}
	\prod_n d\bar{b}'_n da'_n &= \left( \det [M]^{-1} \prod_n d\bar{b}_n \right) \!  \left(\det [M]^{-1} \prod_n da_n \right) \nonumber \\
									&= \left[ \exp \left( -i \sum_n \int d^2x \phi^\dagger_n \alpha\gamma_5\phi_n \right) \prod_n d\bar{b}_n \right] \nonumber \\
									&\times \left[ \exp \left( -i \sum_n \int d^2x \phi^\dagger_n \alpha\gamma_5\phi_n \right)  \prod_n da_n \right] \nonumber \\
									&= \exp \left( -2i \sum_n \int d^2x \phi^\dagger_n \alpha\gamma_5\phi_n \right) \prod_n d\bar{b}_n da_n \nonumber \\
									&\equiv J_5 (\alpha) \prod_n d\bar{b}_n da_n \, ,
	\label{dJacobian}
\end{align}
where $J_5(\alpha)$ is the Jacobian and we have used the identity $\ln \det (M^{-2}) = -2\Tr \ln(M)$ for the second equality. To obtain the corresponding Jacobian for phase rotation \eqref{PhaseTransG}, we can use the prescription above, and observe that its Jacobian is unity and therefore the path-integral measure remains unchanged under phase rotation. This is what we naturally desire, since consequently the particle number conservation remains unbroken at the quantum level. As we can see, chiral rotation has a different story. What was a classical symmetry and generated the continuity equation for chiral current $\partial_\mu j_5^\mu =0$, is now quantum mechanically broken. The broken conservation law now reads:
\begin{equation}
	\int d^2x \partial_\mu \left(\bar{\psi}\gamma^\mu\gamma_5\psi\right)= -2i  \sum_n \int d^2x \phi^\dagger_n \gamma_5\phi_n \, ,
	\label{RawConservation}
\end{equation}
where $\bar{\psi}\gamma^\mu\gamma_5\psi$ is the Noether current $j_5^\mu$ for the classical chiral symmetry.

Looking back at \eqref{MirroredMode} and the statement below it, we find out that only zero modes contribute to the sum in \eqref{RawConservation}, since for all other modes $\int d^2x \phi^\dagger_n \gamma_5\phi_n$ is zero by \eqref{Orthonormality}. The same goes for the exponents of \eqref{dJacobian} as well when $\alpha$ is to be constant. Thus, chiral anomaly is attributed to the subset of zero modes. In fact since zero modes are also eigenvectors of $\gamma_5$, it can be diagonalized in the subspace of zero modes $\gamma_5=\mathrm{diag}(+1,-1)$, and then the value of $\int d^2x \phi^\dagger_n \gamma_5\phi_n$ becomes either $+1$ when $\gamma_5\phi_n=+\phi_n$, or $-1$ when $\gamma_5\phi_n=-\phi_n$. Therefore the sum above is equal to the number of right-moving zero modes $n_+$ minus the number of left-moving zero modes $n_-$,
\begin{equation}
	\sum_n \int d^2x \phi^\dagger_n \gamma_5\phi_n = n_+ - n_- = \mathrm{Index}(\slashed{D}) \, .
	\label{Index}
\end{equation}
Since $\phi_n$s are eigenfunctions of the covariant derivative $\slashed{D}$, they depend on the gauge field $\phi_n \equiv \phi_n(A_\mu(x))$. Therefore, $n_\pm$ also depend on $A_\mu(x)$, which, along with the knowledge that anomaly belongs to zero modes, points to the fact that anomaly is topologically related to the gauge field. It is also worth mentioning that since a zero mode stays a zero mode after a chiral rotation, we can now write \eqref{dJacobian} as
\begin{equation}
	\prod_{n \in \{0\}} d\bar{b}'_n da'_n = e^{-2i\big( n_+[A_\mu] - n_-[A_\mu] \big)} \prod_{n \in \{0\}} d\bar{b}_n da_n \, ,
\end{equation}
where $n$ goes only over the subspace of zero modes, $\{0\}$, while all other modes do not contribute to the Jacobian of transformation.

The dependence of $n_\pm$ on the gauge field is not surprising, as it is clear from the one-dimensional model, applying an electric field, say in the right direction, can of course favor the generation of right moving particles and disfavor the left moving ones. In fact a sober guess can tell us that in (1+1)-dimensional spacetime the number of right moving particles must increase by a rate proportional to $eE$ which is the electrical force exerted on each particle, with $e$ the charge of particles and $E$ the applied electric field. If there was no anomaly, say at the classical level where there is no sea of anti-particles to source this generation, $(n_+ - n_-)$ would have been conserved; instead it is only $(n_+ + n_-)$, the total number of zero modes, that stays conserved at the quantum level. 

We now proceed to calculate what we have guessed above, specifically, how \eqref{Index} is given in terms of the gauge field $A_\mu$. 

\subsection{Regularized Jacobian}

So far, we have found that the Jacobian of chiral transformations $J_5(\alpha)$ is given by what is essentially a trace over $\gamma_5$:
\begin{equation}
	\ln J_5(\alpha) = -2i \lim_{N \to \infty} \sum^N_{n=1} \int d^2x \phi^\dagger_n(x)\alpha(x)\gamma_5\phi_n(x) \, ,
	\label{lnJacobian}
\end{equation}
where we have introduced $N \to \infty$ to emphasize that this sum, roughly speaking, is over an infinite series of $\pm 1$s. It is not trivial that such a sum converges. The final value depends on how we decide to group $+1$s and $-1$s to obtain a convergent series. We need a proper method to sum these numbers which conveys the physics behind. We have already taken one step, namely choosing the basis of the Dirac operator. But the current labels attributed to the eigenfunctions $\phi_n$ in \eqref{lnJacobian} do not carry any physical meaning. The correct method of summation should be gauge invariant so that it can produce a gauge invariant result; otherwise a term resulting from the breaking of gauge symmetry might be confused with the source of anomaly. Therefore, we will further utilize eigenvalues $l_n$ of the Dirac operator, which are invariant under gauge transformations, to relabel the eigenfunctions and regularize \eqref{lnJacobian} as follows.
\begin{align}
	&\frac{i}{2}\ln J_5(\alpha) \nonumber \\
							&= \lim_{M \to \infty}  \int d^2x \alpha(x)  \sum^\infty_{n=1} \phi^\dagger_n(x)\gamma_5 f\left(\frac{l_n^2}{M^2}\right) \phi_n(x) \nonumber \\
							&= \lim_{M \to \infty}  \int d^2x \alpha(x)  \sum^\infty_{n=1} \phi^\dagger_n(x)\gamma_5 f\left(\frac{\slashed{D}^2}{M^2}\right) \phi_n(x) \, ,
	\label{RegularizationIntro}
\end{align}
where $f(x)=1$ for $x<1$ but vanishes for $x>1$ fast enough. Now eigenfunctions are labeled by how large their corresponding eigenvalue is. In \eqref{lnJacobian} we needed to end the summation at some arbitrary $n$ which is reflected in the limit $N\to\infty$, but now we are using $M\to\infty$ and the summation ends on physical grounds.\footnote{At this stage it is worth emphasising on the distinction between the eigen-basis of the Dirac operator, which are used in the regularization, and that of the corresponding Hamiltonian. In particular the zero-modes of the Dirac operator are \textit{not} zero \textit{energy} modes. The zero-modes of $\slashed{D}$ are those which solve the equations of motion of the free theory, in other words the on-shell modes. Therefore, higher eigenvalues belong to those modes which are farther from the shell. This means that the regularization above penalizes the most off-shell degrees of freedom.} Presence of $\slashed{D}$ in the last line of Eq. \eqref{RegularizationIntro} makes it clear how the gauge field $A_\mu$ appears in the anomalous relation. However, what we so far have is given in terms of eigenvalues of $\slashed{D}$ and it is not yet manifest how \eqref{RegularizationIntro} is a function of time and space. To extract the dependence of the regularization $f$ on the coordinates we move to the basis of plane waves.

\begin{subequations}
	\label{2DCal}
\begin{align}
	&\frac{i}{2}\ln J_5(\alpha) = \Tr \alpha\gamma_5 \nonumber \\
	& \equiv \tr \lim_{M \to \infty}  \int \! d^2x \, \alpha \! \int \frac{d^2k}{(2\pi)^2} e^{-ik_\mu x^\mu}\gamma_5 f\left(\frac{\slashed{D}^2}{M^2}\right) e^{ik_\mu x^\mu}  \label{2DCal:a} \\
	&=\tr \lim_{M \to \infty}  \int \! d^2x \, \alpha \! \int \frac{d^2k}{(2\pi)^2} e^{-ik_\mu x^\mu}\gamma_5 f\left(\frac{D^\mu D_\mu -\frac{ie}{4} [\gamma^\mu,\gamma^\nu]F_{\mu\nu}}{M^2}\right) e^{ik_\mu x^\mu}  \label{2DCal:b} \\
	&=\tr \lim_{M \to \infty}  \int \! d^2x \, \alpha \! \int \frac{d^2k}{(2\pi)^2} \gamma_5 f\left(\frac{(-k_\mu k^\mu +2ik^\mu D_\mu + D^\mu D_\mu) -\frac{ie}{4} [\gamma^\mu,\gamma^\nu]F_{\mu\nu}}{M^2}\right)  \label{2DCal:c} \\
	&=\tr \lim_{M \to \infty}  M^2 \! \int \! d^2x\,  \alpha \! \int \frac{d^2k}{(2\pi)^2} \gamma_5 f\left( \! -k_\mu k^\mu +\frac{2ik^\mu D_\mu}{M} + \frac{D^\mu D_\mu}{M^2} -\frac{\frac{ie}{4} [\gamma^\mu,\gamma^\nu]F_{\mu\nu}}{M^2} \! \right) \label{2DCal:d} \\
	&=\tr \lim_{M \to \infty}  - M^2 \int \! d^2x \, \alpha \! \int \frac{d^2k}{(2\pi)^2} f'(-k_\mu k^\mu) \frac{\frac{ie}{4} [\gamma^\mu,\gamma^\nu]\gamma_5 F_{\mu\nu}}{M^2} \label{2DCal:e} \\
	&= i\int \! d^2x \, \alpha \frac{ie}{4\pi}\epsilon^{\mu\nu}F_{\mu\nu} \, . \nonumber
\end{align}
\end{subequations}

In \eqref{2DCal:a} we have chosen the basis of plane waves and have traced over the remaining indices by $\tr$. For the next line, \eqref{2DCal:b}, we have expanded $\slashed{D}^2=\gamma^\mu D_\mu \gamma^\nu D_\nu$ by separating $\gamma^\mu\gamma^\nu$ into its symmetric and anti-symmetric parts and the fact that $[D_\mu,D_\nu] = -ieF_{\mu\nu}$. We have then pulled $e^{ik_\mu x^\mu}$ through covariant derivatives $D_\mu$ from \eqref{2DCal:b} to \eqref{2DCal:c}, which leaves $ik_\mu$ behind wherever there is a covariant derivative $D_\mu \rightarrow ik_\mu + D_\mu$. Changing the variables of integration from $k_\mu$ to $M k_\mu$ leads us to \eqref{2DCal:d}. This will let us expand $f(x)$ in orders of $1/M$ around $-k_\mu k^\mu$ which takes us to the next line. In \eqref{2DCal:e} we have only kept the leading orders of $1/M$ and also have used the fact that all terms coming with only $\gamma_5$ vanish since $\tr \gamma_5 = 0$. Furthermore, in a two-dimensional spacetime $\tr[\gamma^\mu,\gamma^\nu]\gamma_5 = 4 i\epsilon^{\mu\nu}$ which gives us the only surviving term, and we have
\begin{equation}
	\int \frac{d^2k}{(2\pi)^2} f'(-k_\mu k^\mu) = -\int \frac{du}{4\pi} f'(u)= -\frac{1}{4\pi} \, .
\end{equation}
The above equality is true regardless of the details of the function $f$ and is satisfied only by the requirements described before. Therefore, for a general regularizing function we find the Jacobian of chiral transformations for the Minkowski metric to be
\begin{equation}
	J_5 (\alpha) = \exp{ \left\{ -i\int d^2x \alpha(x) \frac{e}{2\pi}\epsilon^{\mu\nu}F_{\mu\nu} \right\} } \, .
\end{equation}
All in all, the differences caused by chiral rotation in the path-integral can be represented by the equality below,
\begin{equation}
	\int \mathcal{D}\bar{\psi}\mathcal{D}\psi e^{iS} = \int \mathcal{D}\bar{\psi}\mathcal{D}\psi e^{iS + i \int \! d^2x \alpha \left( \partial_\mu j_5^\mu - \frac{e}{2\pi}\epsilon^{\mu\nu}F_{\mu\nu} \right) } \, ,
\end{equation}
where from the left-hand side to the right-hand side a chiral rotation in fermionic variables have taken place while the equality sign reflects the fact that the chiral rotation is after all only a change of path-integral variables and must not change the whole path-integral. From the above equality the following anomalous relation is concluded,
\begin{equation}
	\langle \partial_\mu j_5^\mu \rangle = \langle \frac{e}{2\pi}\epsilon^{\mu\nu}F_{\mu\nu} \rangle \, ,
	\label{2DAnomaly}
\end{equation}
where $j_5^\mu \equiv \bar{\psi}\gamma^\mu\gamma_5\psi$ is the Noether current corresponding to chiral rotations and is called ``chiral current''. Had it not been for the non-trivial Jacobian $J_5(\alpha)$, the right hand side of the above equation would have been zero, which is the case for the classical theory were there are no path-integral measures to begin with. Moreover, the angle signs $\langle \, \rangle$ are there to remind us that these equations are path-integral relations.

The chiral anomaly in (1+1) dimensions is probably the simplest form of quantum anomaly, but it is connected to its (3+1) dimensional counterpart if we note that in four dimensions we will have to integrate over a four dimensional momentum space and therefore $M^4$ appears in \eqref{2DCal:d} in place of $M^2$ behind the integral. This means that we now have to expand the regulator $f(x)$ up to the order of $1/M^4$ (instead of $1/M^2$ in two dimensional spacetime), since higher orders in expansion vanish in the limit $M\rightarrow \infty$. On the other hand, algebra of gamma matrices in (3+1) dimensions gives the following
\begin{subequations}
\begin{align}
	\label{trgamma}
	&\tr  \gamma_5 = \tr \left[ \gamma^\mu,\gamma^\nu \right]\gamma_5 = 0 \\
	&\tr \left[ \gamma^\mu,\gamma^\nu \right]\left[ \gamma^\rho,\gamma^\sigma \right] \gamma_5 = -16\epsilon^{\mu\nu\rho\sigma} \, .
\end{align}
\end{subequations}
Thus, the only surviving term from the expansion will come from the square of $\left[ \gamma^\mu,\gamma^\nu \right]F_{\mu\nu}$ which, roughly speaking, gives us the square, $\epsilon^{\mu\nu\rho\sigma}F_{\mu\nu}F_{\rho\sigma}$, of what we had for (1+1) dimensions:
\begin{equation}
	\langle \partial_\mu j^\mu_5 \rangle = \langle \frac{e^2}{16\pi^2}\epsilon^{\mu\nu\rho\sigma}F_{\mu\nu}F_{\rho\sigma} \rangle \, .
	\label{4DAnomaly}
\end{equation}
By following the same reasoning in other dimensions one can show that the chiral anomaly vanishes in odd spacetime dimensions.\footnote{However, it has recently been shown \cite{TopolCriterion} that for certain systems, in particular those which develop localization, it is possible to reduce the dimensionality, by removing the time dimension, from odd to even number of spacetime dimensions hence reviving the chiral anomaly.}
 
 It is also worth while to rewrite \eqref{2DAnomaly} and \eqref{4DAnomaly} in terms of electric and magnetic fields which respectively yields
\begin{equation}
	\langle \partial_\mu j_5^\mu \rangle = \langle \frac{e}{\pi} E  \rangle \, , \quad \quad \text{(2D spacetime)}
	\label{2DAnomalyE}
\end{equation}
and
\begin{equation}
	\langle \partial_\mu j_5^\mu \rangle = \langle \frac{e^2}{2\pi^2}\vec{E} \cdot \vec{B} \rangle \, , \quad \quad \text{(4D spacetime)} 
	\label{4DAnomalyEB}
\end{equation}
where in the latter (four dimensional case) we see that it is the coincidence of electric and magnetic fields that breaks the conservation of the chiral current.

\section{Interaction in 1+1 Dimensions} \label{sec:2D}
We now turn our attention to the following specific question: How will the anomalous relation \eqref{2DAnomaly} be modified, if the fermions are interacting with each other? To this end, consider the interacting action functional below,
\begin{equation}
	S_\lambda = \int d^2x \left[ \bar{\psi} i \gamma^\mu \left( \partial_\mu - ieA_\mu \right) \psi  -\frac{\lambda^2}{2} j^\mu j_\mu \right] \, ,
	\label{2DInetractingAction}
\end{equation}
where $j^\mu \equiv \bar{\psi}\gamma^\mu\psi$ is the Noether current of phase rotation, or simply the electrical current when multiplied by the electrical charge $e$, and  $\lambda^2/2$ is the strength of an attractive interaction. Note that the interaction term does not break any classical symmetry $S_\lambda$ had when $\lambda$ was zero. Knowing this one might readily argue that since the anomaly comes from the measure of the path-integral, and adding a term to the action functional has nothing to do with the measure, then the interaction term $-\frac{\lambda^2}{2}\bar{\psi}\gamma^\mu\psi \bar{\psi}\gamma_\mu\psi$ must have no effect on the anomalous relation whatsoever. This is however \textit{not} the case as we shall now explain.

\subsection{Diagonalized Partition Function}
\label{sec:Regularization}

If the argument in the last sentences above is correct, then interactions must have no effect on the anomalous relation and going any further in this direction would be a lost cause. But from the above argument, the assumption that the action functional has nothing to do with the measure, is wrong.
In this subsection we discuss this fact, and show that it comes from a need for a self-consistent regularization of the path-integral, and postpone a detailed demonstration to appendix \ref{sec:SelfRegularization}.
The reader who is not concerned with such details or would like to return to them later, can safely jump to the next subsection, \ref{sec:Interaction2D}.

Unlike regular path-integration, fermionic path-integrals are defined by left differentiation of Grassmann numbers and they need regularization. For example, path integration over fermionic degrees of freedom of a free Dirac fermion theory yields the determinant of the Dirac operator $\det (i\slashed{D} + m) = \prod l_n$ only if the product is well-defined which is the case when $\sum_n^\infty |l_n -1|$ converges~\cite{InfiniteProduct}. But eigenvalues, $l_n$, of the Dirac operator are not even bounded. Thus the well-defined path-integral must carry a type of regularization with itself. This is where action functional intrudes into measure's business. The path-integral (or the partition function) has two elements; the measure and the action; the regularization is not present in the latter. But \textit{how} the measure is regularized must only be determined by the action, since the partition function is to be a self-sufficient object and should be indifferent about the backstory of why it is written. Let us again look at the free Dirac fermion. We can write its partition function as,
\begin{align}
    \det\left(i\slashed{D}+m\right) &= \int \prod_n d\bar{b}_n d a_n \left( 1 + l_n \bar{b}_n a_n \right)  = \int \left( \prod_n  d\bar{b}_n da_n\right) \exp\left\{\sum_n l_n \bar{b}_n a_n\right\} \, ,
\end{align}
with $\bar{b}_n$ and $a_n$ being the Grassmann amplitudes defined before in equation \eqref{SpinorExpansion}. Looking at above we see that the action is given by $\sum_n l_n \bar{b}_n a_n$. This setup, in which the action is \textit{formally diagonalized}, makes it easy for us to set a natural cut-off on the determinant: We can simply disregard all $l_n$s which are bigger than some limit $M$ and eventually take the limit $M \rightarrow \infty$. This is exactly what we have done for calculating the chiral anomaly in the previous section.

The above way of regularizing the path-integral clearly relies on the action, since the cut-off is actually on the eigenvalues of the Euclidean action. In other words, the basis chosen for the spinor expansion of the fermionic degrees of freedom, are those which formally diagonalize the action functional. After this diagonalization the regularization is naturally given as above. Note also that this regularization procedure is indifferent to external information such as what physical system is being described by the path-integral, or whether a particular field has a certain symmetry or not. The regularization, being determined only by the action, will share the symmetries of the action; even if the measure does not respect them all.

Now we can go back to question raised below equation \eqref{2DInetractingAction}. To calculate the anomaly we first have to expand the fermionic degrees of freedom in spinor modes. But what basis should they be expanded in? The answer would be the basis that formally diagonalizes the action and therefore leads to a well-defined path-integral. Clearly, by introducing interaction terms to the action, the basis in which it will be diagonalized will change accordingly. Through this, the presence of interactions will modify the anomaly.

This subject is rigorously expounded in the appendix \ref{sec:SelfRegularization} although the material covered there will not be essential for what follows.

\subsection{Effect of Interaction}
\label{sec:Interaction2D}

Now we can proceed to see how the interactions modify the anomalous relation in $(1+1)$ dimensions. In the case of \eqref{2DInetractingAction} we can easily find a regularization that fits the descriptions put forward in the previous subsection and in appendix \ref{sec:SelfRegularization}, although this regularization, which follows shortly, may be natural enough for the reader to ignore the previous subsection altogether.

We can decouple the interaction term using a Hubbard-Stratonovich auxiliary field, $a_\mu$ (not to be confused with the Grassmann variable $a_n$), arriving at the following equality
\begin{equation}
	\int \mathcal{D}\bar{\psi}\mathcal{D}\psi e^{S_\lambda} = \int \mathcal{D}\bar{\psi}\mathcal{D}\psi \mathcal{D}a_\mu e^{S_a} \, ,
	\label{IntHS}
\end{equation}
with $S_\lambda$ given by \eqref{2DInetractingAction} and $S_a$ as follows,
\begin{equation}
	S_a = \int d^2x  \left[ \bar{\psi} i \gamma^\mu \left( \partial_\mu - ieA_\mu -i\lambda a_\mu \right) \psi  + \frac{1}{2}a_\mu a^\mu \right] \, .
	\label{2DHSAction}
\end{equation}
To see the equality of the two path-integrals, we first shift $a_\mu$ in \eqref{2DHSAction} by $-\lambda \bar{\psi}\gamma_\mu\psi$ to obtain
\begin{equation}
	S_a = \int d^2x  \left[ \bar{\psi} i \slashed{D} \psi  -\frac{\lambda^2}{2} j_\mu j^\mu + \frac{1}{2}a_\mu a^\mu \right] \, ,
\end{equation}
but a simple shift does not change the path-integral measure $\mathcal{D}a_\mu$. We can then integrate $a_\mu$ out and regain equation \eqref{2DInetractingAction}. Recall that path-integral equalities of the form \eqref{IntHS} ought to be regarded as equalities upto a constant coefficient.

What has been sandwiched between $\bar{\psi}$ and $\psi$ is now a generalized Dirac operator $\slashed{D}_g \equiv \gamma^\mu \left( \partial_\mu - ieA_\mu - i\lambda a_\mu \right)$ where $A_\mu$ and $a_\mu$ have the same mathematical status. It is therefore rational to think that $\slashed{D}_g$ must substitute $\slashed{D}$ in Fujikawa's method of regularization introduced in \eqref{RegularizationIntro}. Regardless of naturalness, moreover, regularizing with respect to $\slashed{D}_g$ satisfies the requirements put forward in the previous section and appendix \ref{sec:SelfRegularization}, namely expanding the spinor degrees of freedom in the basis of $\slashed{D}_g$ formally diagonalizes the fermionic part of the action.\footnote{To make a connection to what has been suggested in appendix \ref{sec:SelfRegularization} one can look at \eqref{EigenIntegral}, \eqref{RegularizationBasis} and \eqref{FormallyDiagonalized}. The fact that such a regularization is enforced can be observed simply by choosing the basis of generalized hermitian Dirac operator $\slashed{D}_g \phi_n = l_n \phi_n$ in order to expand fermionic degrees of freedom $\bar{\psi}$ and $\psi$ as in \eqref{SpinorExpansion}, which makes the fermionic part of the action $S_a$ formally diagonalized:
\begin{align}
	I = \int &\mathcal{D}a_\mu \exp \left\{ \frac{1}{2}\int d^2x a_\mu a^\mu \right\} \times \int \prod_n d\bar{b}_n a_n \exp\left\{ \sum_n l_n \bar{b}_n a_n \right\} \, .
\end{align}
Then, for instance, equation \eqref{EigenIntegral} tells us that $\ell_n = l_n$, meaning that the eigenvalues of the generalized Dirac operator $\slashed{D}_g$ determine the regularization.

It is also worth noting that the on-shell modes are now the zero-modes of the generalized Dirac operator, $\slashed{D}_g \phi_{\{0\}} = 0$. So the regularization process should assign a penalty to the off-shell modes with respect to this operator.}

Having established the above we can proceed to calculate the anomalous relation in presence of interactions. We need to calculate $\sum_n \phi^\dagger \gamma_5 f(\slashed{D}_g^2/M^2) \phi_n$ where we can take the same steps as \eqref{2DCal} but with $eA_\mu$ replaced by $eA_\mu + \lambda a_\mu$, which yields,
\begin{equation}
	\langle \partial_\mu j_5^\mu \rangle_{a_\mu} = \langle \frac{e}{\pi} \epsilon^{\mu\nu} \partial_\mu A_\nu + \frac{\lambda}{\pi}\epsilon^{\mu\nu}\partial_\mu a_\nu \rangle _{a_\mu} \, ,
\end{equation}
where the subscript $a_\mu$ is a reminder that in addition to fermionic fields there is also an integration over the auxiliary field $a_\mu$. We have also used the following,
\begin{equation}
	\epsilon^{\mu\nu} F_{\mu\nu} = \epsilon^{\mu\nu} \partial_\mu A_\nu + \epsilon^{\nu\mu}  \partial_\nu A_\mu = 2\epsilon^{\mu\nu}\partial_\mu A_\nu \, .
\end{equation}
To get back to the original path-integral $\int\mathcal{D}[\bar{\psi},\psi] e^{S_\lambda}$, we must first shift $a_\mu$ to $a_\mu - \lambda j_\mu$ and then integrate it out,
\begin{align}
	& \langle \partial_\mu j_5^\mu \rangle_{a_\mu} = \langle \frac{e}{\pi} \epsilon^{\mu\nu} \partial_\mu A_\nu -\frac{\lambda^2}{\pi}\epsilon^{\mu\nu}\partial_\mu j_\nu  + \frac{\lambda}{\pi}\epsilon^{\mu\nu}\partial_\mu a_\nu \rangle _{a_\mu}  \nonumber \\
		& \Rightarrow \partial_\mu j_5^\mu = \frac{e}{\pi} \epsilon^{\mu\nu} \partial_\mu A_\nu -\frac{\lambda^2}{\pi}\epsilon^{\mu\nu}\partial_\mu j_\nu \, ,
\end{align}
where in the last line an integration over fermionic degrees of freedom is implied. Note that the action is even in $a_\mu$, therefore, odd terms such as $\langle a_\mu \rangle$ are eliminated after the integration.

We can look at the result in different ways one of which is to write in the following form,
\begin{equation}
	\partial_\mu j_5^\mu = \frac{e}{\pi}\epsilon^{\mu\nu} \partial_\mu \left( A_\nu +\frac{\lambda^2}{e^2} ej_\nu \right) \, ,
	\label{2DIntScreen}
\end{equation}
and see the interplay of the interaction and the chiral anomaly manifested as an addition to the electromagnetic field due to the electrical current $ej^\mu$. Note that the anomalous term above is present also in the absence of the gauge field; its effect then is to renormalize the excitations of the system. To see this recall that in (1+1)-dimensions we have $\gamma^\mu\gamma_5 = \epsilon^{\mu\nu}\gamma_\nu$, allowing us to write $j^\mu_5 = \epsilon^{\mu\nu}j_\nu$, and therefore, restricted to (1+1)-dimensions, write the anomalous relation as,
\begin{equation}
	\partial_\mu j_5^\mu = \frac{1}{1+\lambda^2/\pi} \frac{e}{2\pi}\epsilon^{\mu\nu}F_{\mu\nu} \, ,
	\label{2DIntAnomaly}
\end{equation}
which gives us the same anomalous term on the right hand side as in the absence of interactions but now modified by a coefficient $(1+\lambda^2/\pi)^{-1}$.

For an effective field theory of a condensed matter system there is no real reason for preserving Lorentz symmetry. Thus it is reasonable to investigate interactions such as the density-density interaction, $(\bar{\psi}^\dagger\psi)^2$, that do not respect it. This type of interaction completely fits to our procedure if we substitute $a_\mu$ just by its temporal component $\delta^0_\mu a_0$ and integrate only over $a_0$ instead. We get
\begin{equation}
	\partial_\mu j_5^\mu = \frac{e}{2\pi}\epsilon^{\mu\nu}F_{\mu\nu} - \frac{\lambda^2}{\pi}\partial_1 j_5^1 \, ,
\end{equation}
which shows that even in the absence of external electromagnetic fields, the chiral charge conservation law is modified by the interplay of interactions and chiral anomaly. In the absence of the external field the anomalous terms still remain and their effect can be seen as a renormalization of the velocity of the excitations which carry the chiral charge.

\subsection{Finite Chiral Rotation}

All the transformations we were concerned with so far were infinitesimal transformations, allowing us to obtain a (non)-conservation relation. In this subsection we will see how finite transformations differ with infinitesimal ones in the context of anomalies.

An infinitesimal chiral rotation $\psi \rightarrow e^{i\alpha\gamma_5}\psi$ adds a term of $ \gamma^\mu (i\partial_\mu \alpha) \gamma_5$ to the Dirac operator $\slashed{D}$ and an anomalous term, $-\alpha\frac{e}{2\pi}\epsilon^{\mu\nu}F_{\mu\nu}$, coming from the non-trivial Jacobian, to the Lagrangian. However, if you consider the effect of the former on the anomaly, you can observe that the anomalous term should have been modified, but since $\alpha$ is infinitesimal, the higher order corrections can be disregarded. But a finite chiral rotation as it appears inside the generalized Dirac operator $\slashed{D}_g$, changes the anomalous term and its potentially sizable contributions must be brought into account.

We begin by noticing that in two dimensional spacetime $\gamma^\mu\gamma_5$ is equal to $\epsilon^\mu_{\ \nu}\gamma^\nu$ so that we are able to write a generalized Dirac operator which carries an axial-vector term as
\begin{align}
	\slashed{D}_g &= \gamma^\mu \left( \partial_\mu -ieA_\mu -ib_\mu\gamma_5 \right) \nonumber \\
		&= \gamma^\mu \left( \partial_\mu -ieA_\mu - i\epsilon^\nu_{\ \mu} b_\nu \right) \equiv \gamma^\mu \left( \partial_\mu -iC_\mu \right) \, .
\end{align}
with $b_\mu$ being the axial-vector field. And therefore its corresponding anomalous relation is
\begin{equation}
	\partial_\mu j^\mu_5 = \frac{1}{\pi}\epsilon^{\mu\nu}\partial_\mu C_\nu = \frac{e}{\pi}\epsilon^{\mu\nu}\partial_\mu A_\nu + \frac{1}{\pi}\partial_\mu b^\mu \, ,
\end{equation}
where the right hand side is due to the Jacobian in presence of $b_\mu$. We see that if $b_\mu$ is a constant across spacetime it will not contribute.

Now let us start over with a simple Dirac operator $\slashed{D} \equiv \gamma^\mu (\partial_\mu - ieA_\mu)$, and arrive at a finite chiral rotation in steps of $d\alpha$ so that at any certain step an angle of $\alpha=\int d\alpha$ has been accumulated, which finally reaches $\alpha_f$. At each step the following term is added to the action
\begin{align}
	\delta_{d\alpha} S = \int d^2x \bigg[ &-(\partial_\mu d\alpha)j_5^\mu - d\alpha \frac{e}{\pi}\epsilon^{\mu\nu} \partial_\mu A_\nu + \frac{d\alpha}{\pi}\partial_\mu \partial^\mu \alpha \bigg] \, .
\end{align}
Hence by integrating $d\alpha$ the total change due to the finite chiral rotation is obtained to be,
\begin{equation}
	\delta_\alpha S = \int d^2x \left[ -b_\mu j_5^\mu + \frac{e}{\pi}\epsilon^{\mu\nu} b_\mu A_\nu + \frac{1}{2\pi}\partial_\mu b^\mu \right] \, ,
\end{equation}
with $b_\mu$ set equal to $\partial_\mu \alpha_f$. One conclusion here is that if we begin with the action below,
\begin{equation}
	S = \int d^2x \bar{\psi}i\gamma^\mu \left( \partial_\mu - ieA_\mu - ib_\mu \gamma_5 \right) \psi \, ,
\end{equation}
with $b_\mu$ constant, we can remove $b_\mu$ from the Dirac operator by a chiral rotation with angle $\alpha$ satisfying $\partial_\mu \alpha = b_\mu$.
\begin{equation}
	S + \delta_\alpha S = \! \int \! d^2x \bigg[ \bar{\psi}i\gamma^\mu \left( \partial_\mu - ieA_\mu \right) \psi + \frac{e}{\pi}\epsilon^{\mu\nu} b_\mu A_\nu   \bigg] \, .
\end{equation}

\subsection{Effective Action} \label{sec:EffAct}

In 2D vector spaces, Helmholtz theorem takes quite a simple form and allows us to decompose any $\slashed{v}$ into vector and axial-vector parts, $\slashed{v}= \slashed{\partial} \rho_v + \slashed{\partial} \phi_v \gamma_5$, with $v_\mu$ being a two-vector and $\rho_v$ and $\phi_v$ two scalars, generating the divergence-free and curl-free parts of the vector field $v_\mu$ respectively. Returning to the (1+1)-dimensional Hubbard-Stratonovich action functional \eqref{2DHSAction},
\begin{equation}
	S_a \! = \! \int \! \! d^2x \! \left\{ \bar{\psi}i\gamma^\mu\left[ \partial_\mu -ieA_\mu -i\lambda a_\mu \right] \psi +\frac{1}{2}a_\mu a^\mu \right\} \, ,
\end{equation}
we can rewrite $A_\mu$ and $a_\mu$ as the sum of their curl-free and divergence-free parts,
\begin{align}
	S_a = \! \int \! d^2x  \bigg\{ &\bar{\psi}i\gamma^\mu\left[ \partial_\mu -ie\epsilon_\mu^{\ \nu}\partial_\nu\Phi -i\lambda (\partial_\mu \rho + \epsilon_\mu^{\ \nu}\partial_\nu\phi) \right] \psi + \frac{1}{2}\partial_\mu\rho\partial^\mu \rho - \frac{1}{2}\partial_\mu\phi\partial^\mu\phi \bigg\}.
\end{align}
where we have chosen the Lorentz gauge $\partial_\mu A^\mu = \partial_\mu \partial_\nu \left( \epsilon^{\mu \nu} \Phi \right)=0$ and used the identity $\epsilon_{\mu\alpha}\epsilon^\mu_{\ \beta}=-\eta_{\alpha\beta}$. In two space-time dimensions we can write $\gamma^\mu\gamma_5=\epsilon^\mu_{\ \nu}\gamma^\nu$ and rearrange the above equation into
\begin{align}
	S_a = \int d^2x  \bigg\{ &\bar{\psi}i\gamma^\mu\left[ \partial_\mu + i\gamma_5 (e\partial_\mu\Phi +\lambda\partial_\mu\phi) -i\lambda\partial_\mu \rho \right] \psi  + \frac{1}{2}\partial_\mu\rho\partial^\mu \rho - \frac{1}{2}\partial_\mu\phi\partial^\mu\phi \bigg\} \, .
\end{align}
Next we are going to draw out the $\gamma_5$ from inside the brackets by a finite chiral rotation $\psi \rightarrow e^{-i(e\Phi + \lambda\phi)\gamma_5}\psi'$. This in turn introduces a Jacobian which appears in the action as
\begin{align}
	S_a &=  \int d^2x  \bigg\{ \bar{\psi}'i\gamma^\mu\partial_\mu \psi' + \frac{1}{2}\partial_\mu\rho\partial^\mu \rho \nonumber \\
	& -\frac{1}{2\pi}\partial_\mu(e\Phi + \lambda\phi) \partial^\mu(e\Phi +\lambda\phi) - \frac{1}{2}\partial_\mu\phi\partial^\mu\phi \bigg\} \, ,
\end{align}
where we have also eliminated $-i\lambda\partial_\mu \rho$ by a $U(1)$ phase rotation without any cost. A simple squaring process gives
\begin{eqnarray} \label{BeforeInt}
	S_a = \int d^2x &&\bigg\{  \bar{\psi}'i\gamma^\mu\partial_\mu \psi' + \frac{1}{2}\partial_\mu\rho\partial^\mu \rho \nonumber \\
		&&- \frac{1+\lambda^2/\pi}{2} \left(\partial_\mu\phi + \frac{\lambda e/\pi}{1+\lambda^2/\pi} \partial_\mu\Phi \right)^2 -\frac{1}{1+\lambda^2/\pi} \frac{e^2}{2\pi}\partial_\mu\Phi\partial^\mu\Phi \bigg\} \, .
\end{eqnarray}
Here we can easily shift $\phi$ so that it devours the other $\Phi$ term inside the parentheses, and then we can integrate $\phi$, $\rho$ and fermionic degrees of freedom out, to end up with,
\begin{equation} \label{ModifiedMass}
	S \equiv S[A^\mu] = \int d^2x  \, \frac{1}{1+\lambda^2/\pi} \frac{e^2}{2\pi}A_\mu A^\mu \, ,
\end{equation}
where we remembered that $A^\mu = \epsilon^{\mu\nu}\partial_\nu\Phi$ whenever $\partial_\mu A^\mu =0$. As we can see the effective field theory for the gauge field in (1+1)-dimensions comes with a mass term for photon which is modified by interactions. Moreover, by functionally differentiating \eqref{2DHSAction} and \eqref{ModifiedMass} with respect to $eA_\mu (x)$ we find
\begin{equation}
	j^\mu = \frac{1}{1+\lambda^2/\pi} \frac{e}{\pi} A^\mu \, .
\end{equation}
Notice, however, that this equation is true if we begin by choosing the Lorentz gauge for $A_\mu$, this consequently means that electrical current $j^\mu$ is conserved. On the other, hand chiral current $j^\mu_5 = \epsilon^\mu_{\ \nu} j^\nu$ is not and its non-conservation is modified by interaction strength $\lambda^2$.

Getting back to \eqref{BeforeInt} let us only integrate the auxiliary $\rho$ and $\phi$ fields out and leave the Grassmann fields unintegrated. We will have,
\begin{align}
	S = \int d^2x \bigg\{  &\bar{\psi}'i\gamma^\mu\partial_\mu \psi'  - \frac{1}{2\pi}\partial_\mu\frac{e\Phi}{\sqrt{1+\lambda^2/\pi}} \partial^\mu \frac{e\Phi}{\sqrt{1+\lambda^2/\pi}} \bigg\} \, .
\end{align}
By another (reverse) rotation $\psi' \rightarrow e^{-i\gamma_5 e\Phi/\sqrt{1+\lambda^2/\pi}}\Psi$ we can now re-couple the gauge field to fermion number current. This gives us
\begin{equation}
	S = \int d^2x \, \bar{\Psi}i\gamma^\mu \left[ \partial_\mu -i\frac{e}{\sqrt{1+\lambda^2/\pi}}\gamma_5 \partial_\mu \Phi \right] \Psi \, .
\end{equation}
Using the relation $\gamma^\mu\gamma_5=\epsilon^\mu_{\ \nu}\gamma^\nu$ once again, we can write the above as
\begin{equation}
	S = \int d^2x \, \bar{\Psi}i\gamma^\mu \left[ \partial_\mu -i\frac{e}{\sqrt{1+\lambda^2/\pi}}A_\mu \right]  \Psi \, .
\end{equation}

This means that in (1+1)-dimensions the effect of the current-current interaction is summarized in having a \textit{free} fermionic theory with a ``dressed'' charge of $e/\sqrt{1+\lambda^2/\pi}$. 

\section{Interaction in 3+1 Dimensions} \label{sec:4D}

Having the basics established, we are going to continue, for the case of four spacetime dimensions, with a more general short range current-current interaction appearing in the following path-integral as,
\begin{equation}
		I=\int \! \mathcal{D}[\bar{\psi}\psi] \exp \left\{ i \int \! d^4x \left[ \bar{\psi}i\slashed{D}\psi - \frac{1}{2} \lambda^2_{\mu\nu} j^\mu j^\nu \right] \right\} ,
		\label{InteractingI}
\end{equation}
where the current $j^\mu$ and the Dirac operator $\slashed{D}$ are defined as before and $\lambda^2_{\mu\nu} \equiv \lambda_{\mu\alpha}\lambda_\nu^{\ \alpha}$ is the interaction strength. 
The methods we outline in this paper are quite general and can be applied to arbitrary interaction strengths, however for clarity, at times, we have restricted our focus to special cases of $\lambda^2_{\mu\nu}=\lambda^2\eta_{\mu\nu}$, which preserves Lorentz symmetry and $\lambda^2_{\mu\nu}=\lambda^2_0\eta_{0\mu}\eta_{0\nu} + \lambda^2_3\eta_{3\mu}\eta_{3\nu}$. In the latter, when $\lambda_3^2=0$ the interaction term simply becomes the density-density interaction; on the other hand, when $\lambda_3^2=\lambda^2_0$, it breaks the full Lorentz symmetry down to a reduced symmetry constructed by a rotational invariance in the $x$-$y$ plane and a boost invariance along the $z$ direction. The same symmetry reduction happens in the presence of a constant background magnetic field along the longitudinal direction, $z$.
Evidently, depending on the choice of $\lambda_{\mu\nu}$ some of the symmetries of the model may be broken, e.g. Lorentz invariance, but they do not break the classical chiral symmetry. Unlike in $(1+1)$ dimensions, these interactions are RG irrelevant and typically are not considered, however we will see that in the presence of the constant magnetic field, they should not be discounted.

\subsection{Interacting Anomalous Relation}

Again, through Hubbard-Stratonovich decoupling the path-integral $I$ can also be written as
\begin{equation}
		I=\int \! \mathcal{D}[\bar{\psi}\psi a_\mu] \exp  \left\{ i \! \int \! d^4x \left[ \bar{\psi}i\slashed{D}_g\psi +\frac{1}{2}a_\mu a^\mu \right] \right\}
		\label{DecoupledI} ,
\end{equation}
with the generalized Dirac operator $\slashed{D}_g$ now defined as $\slashed{D}_g \equiv \gamma^\mu\left( \partial_\mu -ieA_\mu - i\lambda_{\mu\nu}a^\nu \right)$. The equivalence of \eqref{InteractingI} and \eqref{DecoupledI} can be observed by noticing that a shift in the auxiliary field by its on-shell value $a_\mu \rightarrow a_\mu - \lambda_{\nu\mu}j^\nu$ in the above equation, preserves the measure but changes the action to that of \eqref{InteractingI} plus a quadratic term in $a_\mu$ which can be integrated out leaving only a constant. Therefore these two path-integrals and all correlation functions generated by them are concluded to be equivalent. In what follows, we are going to reserve $S_\lambda$ for the action of \eqref{InteractingI} and $S_a$ for the action of \eqref{DecoupledI} as in the two-dimensional case.

An infinitesimal chiral rotation $\bar{\psi} \rightarrow \bar{\psi}e^{i\alpha\gamma_5}$, $\psi \rightarrow e^{i\alpha\gamma_5} \psi$, transforms the action $S_a$ to 
\begin{equation}
		S_a + \delta S_a \equiv S_a +  \int d^4x \, \alpha (\partial_\mu j_5^\mu - \mathcal{A}_5) \, .
\end{equation}
The first term in the parentheses arises from the classical shift of the action itself whereas the second is the anomalous term introduced by the non-invariance of the measure, or in other words, the non-trivial Jacobian of the transformation. From the path-integral point of view, the transformation is a mere renaming of the variables of integration. Demanding the same value for the path-integral after such change of variables, is now translated in requiring $\delta S_a$ to vanish. The pre-regularization anomalous term is given similar to its two-dimensional counterpart as, 
\begin{equation}
	\int \! d^4x \, \mathcal{A}_5 (x) \equiv \! \int \! d^4x \left[ 2\lim_{N\rightarrow\infty} \sum^N_{n=1}\phi^\dagger_n (x) \gamma_5 \phi_n (x) \right] .
	\label{A5}
\end{equation}
As we discussed in sections \ref{sec:Regularization} and \ref{sec:Interaction2D}, the above is well-defined after a regulator $f(l_n^2/M^2)$ is introduced in the sum, where $l_n$s are the eigenvalues of the generalized Dirac operator $\slashed{D}_g$ while $\phi_n$s are the corresponding eigenfunctions. Then the large value limit of $N$ gives way to that of $M$ as in \eqref{RegularizationIntro},
\begin{equation}
	\mathcal{A}_5 (x)= \lim_{M \to \infty}  \sum^\infty_{n=1} \phi^\dagger_n(x)\gamma_5 f\left(\frac{\slashed{D}_g^2}{M^2}\right) \phi_n(x) \, ,
	\label{RegA5}
\end{equation}
where now $\slashed{D}_g$ is able to capture the position dependence of anomaly after acting on $\phi_n(x)$s which can further be replaced by plane waves via a change of basis. The four dimensional counterpart of the calculation done in \eqref{2DCal}, as was discussed above equation \eqref{4DAnomaly}, determines the value of $\mathcal{A}_5$ to be,
\begin{align}
	\mathcal{A}_5 = \frac{e^2}{16\pi^2} \epsilon^{\mu\nu\rho\sigma} \mathcal{F}_{\mu\nu}\mathcal{F}_{\rho\sigma} \, ,
\end{align}
with $\mathcal{F}_{\mu\nu} \equiv \partial_\mu (eA_\nu+\lambda_{\nu\alpha}a^\alpha)-\partial_\nu (eA_\mu+\lambda_{\mu\beta}a^\beta)$ which renders the anomalous non-conservation law as below,
\begin{align} 	\label{4DMidInt}
	\left\langle \partial_\mu j^\mu_5 \right\rangle_a &= \frac{e^2}{16\pi^2} \left\langle  \epsilon^{\mu\nu\rho\sigma} \mathcal{F}_{\mu\nu}\mathcal{F}_{\rho\sigma} \right\rangle_a \\
	& = \frac{e^2}{16\pi^2} \left\langle \epsilon^{\mu\nu\rho\sigma} F_{\mu\nu} F_{\rho\sigma} \right\rangle_a   \nonumber \\ 
	 & \quad \  + \frac{e}{2\pi^2}\epsilon^{\mu\nu\rho\sigma}   \left\langle \partial_\mu A_\nu \partial_\rho  (\lambda_{\sigma\alpha} a^\alpha) \right\rangle_a 
	 +  \frac{1}{4\pi^2}\epsilon^{\mu\nu\rho\sigma} \left\langle \partial_\mu (\lambda_{\nu\alpha} a^\alpha)  \partial_\rho (\lambda_{\sigma\beta} a^\beta)  \right\rangle_a \, , \nonumber
\end{align}
where we remember that all relations here are path-integral relations and in particular there is an integral over the auxiliary field $a_\mu$ in the above which establishes the path-integral equivalence
\begin{equation}
	\int \mathcal{D}\bar{\psi}\mathcal{D}\psi e^{S_\lambda} = \int \mathcal{D}\bar{\psi}\mathcal{D}\psi \mathcal{D}a_\mu e^{S_a} \, .
\end{equation}

We want to integrate $a_\mu$ out in equation \eqref{4DMidInt} and get the original path-integral \eqref{InteractingI} with the current-current interaction term back. To go from RHS to LHS in the above relation, we first shift the Hubbard-Stratonovich field by its on-shell value $a_\mu \rightarrow a_\mu - \lambda_{\nu\mu}j^\nu$ which turns \eqref{4DMidInt} into,
\begin{align}
	\left\langle \partial_\mu j^\mu_5 \right\rangle_a  &= \frac{e^2}{16\pi^2} \left\langle \epsilon^{\mu\nu\rho\sigma} F_{\mu\nu} F_{\rho\sigma} \right\rangle_a   \\
	&  + \frac{e}{2\pi^2}\epsilon^{\mu\nu\rho\sigma}  \left\langle \partial_\mu A_\nu \partial_\rho  (\lambda_{\sigma\alpha} a^\alpha - \lambda^2_{\sigma\alpha}j^\alpha) \right\rangle_a  \nonumber \\ 
	&  + \frac{1}{4\pi^2}\epsilon^{\mu\nu\rho\sigma} \left\langle \partial_\mu (\lambda_{\nu\alpha} a^\alpha - \lambda^2_{\nu\alpha}j^\alpha)  \partial_\rho (\lambda_{\sigma\beta} a^\beta - \lambda^2_{\sigma\beta}j^\beta)  \right\rangle_a \, . \nonumber
\end{align}
In the above all terms are odd in $a_\mu$ except the term $\epsilon^{\mu\nu\rho\sigma}\lambda_\nu^{\ \alpha}\lambda_\sigma^{\ \beta}\partial_\mu a_\alpha \partial_\rho a_\beta$ coming from the last line. But if the matrix $\lambda_\mu^{\ \nu}$ has only one non-zero element in each of its rows and columns, which is the case for all special cases that we are focusing on, then this term is also odd in each component of $a_\mu$. The reason is that the totally anti-symmetric tensor $\epsilon^{\mu\nu\rho\sigma}$ does not allow for repeated indices. Therefore, after integrating the auxiliary field $a_\mu$ out, all terms that contain $a_\mu$ will vanish, and we will be left with,
\begin{align}\label{FullWard}
	\partial_\mu j^\mu_5 &=\frac{e^2}{16\pi^2} \epsilon^{\mu\nu\rho\sigma} F_{\mu \nu}F_{\rho \sigma}
	-\frac{e}{2\pi^2}\epsilon^{\mu\nu\rho\sigma}\lambda^2_{\sigma\alpha}\partial_\mu A_\nu\partial_\rho j^\alpha +\frac{1}{4\pi^2}\epsilon^{\mu\nu\rho\sigma}\lambda^2_{\nu\alpha}\lambda^2_{\sigma\beta}\partial_\mu j^\alpha \partial_\rho j^\beta \, ,
\end{align}
We see that there are terms respectively depending only on the electromagnetic field, only on the presence of interactions, and a mixed term requiring the presence of both. Note that the first term on the right hand side is the anomalous term in the absence of interactions.

\subsection{Interpretation through Screening}

It is possible to look at the above result as a screening process which is the subject of this subsection.
The anomalous identity in the absence of interactions was previously given by \eqref{4DAnomalyEB} which is $\partial_\mu j^\mu_5=\frac{e^2}{2\pi^2} \vec{E}\cdot\vec{B}$. In comparison, by defining
\begin{eqnarray}\label{D}
	\tilde{E}_i&=&E_i-\frac{1}{e}\left[\lambda^2_{i\beta}\partial_0-\lambda^2_{0\beta}\partial_i\right]j^\beta \, ,\\\label{H}
	\tilde{B}_i&=&B_i-\frac{1}{2e}\epsilon_{ijk}\left[\lambda^2_{j\beta}\partial_k-\lambda^2_{k\beta}\partial_j\right]j^\beta \, ,
\end{eqnarray}
the anomalous identity in presence of interactions \eqref{FullWard} can also take the following appearance
\begin{equation}
	\partial_\mu j^\mu_5 = \frac{e^2}{2\pi^2}\tilde{\vec{E}} \cdot \tilde{\vec{B}} \, ,
\end{equation}
juxtaposed to the (1+1)-dimensional counterpart, equation \eqref{2DIntScreen}. Viewed this way, one can interpret the effects of interaction as a sort of anomalous screening where the newly introduced interactions between charge densities and currents result in the screening of constituents of the external electromagnetic field responsible for the chiral symmetry breaking.

The interacting identity above can equivalently be written as,
\begin{equation}
	\partial_\mu j^\mu_5 = \frac{e^2}{16\pi^2}\epsilon^{\mu\nu\rho\sigma} \tilde{F}_{\mu\nu} \tilde{F}_{\rho\sigma} \, ,
	\label{tildeF}
\end{equation}
where $\tilde{F}_{\mu\nu} \equiv F_{\mu\nu} - \frac{1}{e}\left[ \partial_\mu \left(\lambda^2_{\nu\alpha}j^\alpha \right) - \partial_\nu \left(\lambda^2_{\mu\alpha}j^\alpha \right) \right]$ with the brackets being the contribution from the interacting currents.
Let us for the moment consider the electromagnetic field to be dynamical and focus on the case $\lambda^2_{\mu\nu}=\lambda^2 \eta_{\mu\nu}$ where interactions respect Lorentz symmetry. Upon treating the electromagnetic field in a semi-classical fashion i.e. saddle-point approximation of $A_\mu$ while it is coupled to fluctuating fermionic fields $\psi$ and $\bar{\psi}$, we have $\partial_\nu F^{\mu\nu}=ej^\mu$ and consequently $\partial_\mu j^\mu = 0$. Therefore, Maxwell's equations for $\tilde{F}_{\mu\nu}$ are given by the four components of
\begin{equation}
\partial_\nu \tilde{F}^{\mu\nu} = \left(1-\frac{\lambda^2}{e^2}\Box \right) ej^\mu \, .
\end{equation}
On the other hand, using the commutative nature of partial derivatives, we have
\begin{equation}
	\tilde{F}_{\mu\nu}= \left( 1 - \frac{\lambda^2}{e^2}\Box \right) F_{\mu\nu} \, ,
\end{equation}
or equivalently, 
\begin{equation}
	\tilde{A}_\mu = \left( 1 - \frac{\lambda^2}{e^2}\Box \right) A_\mu \, ,
\end{equation}
where $\tilde{A}_\mu$ is defined through $\tilde{F}_{\mu\nu} \equiv \partial_\mu \tilde{A}_\nu - \partial_\nu \tilde{A}_\mu$. Therefore, looking both at equation \eqref{tildeF} and equations above we see that the anomalous chiral symmetry breaking is generated not only by the background fields but also by backreactions of the interacting matter.

\subsection{A Perturbative Discussion}

\begin{figure}
    \includegraphics[width=\linewidth]{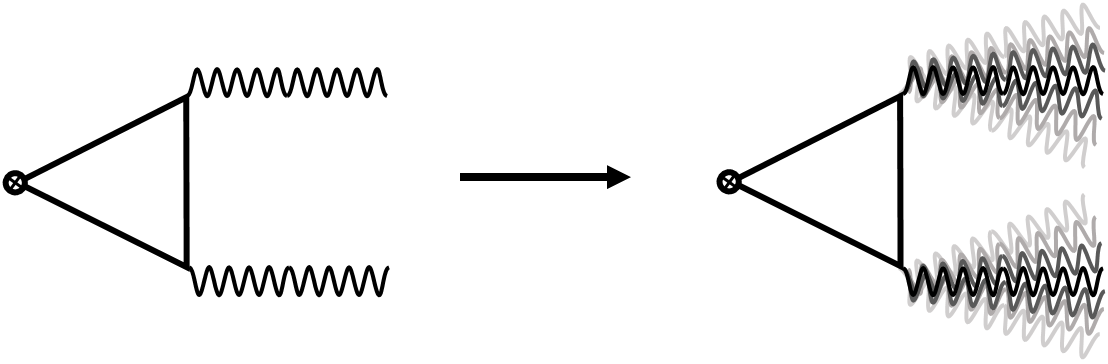}
    \centering
    \caption{Chiral anomaly is associated to triangle diagrams. On the left: The triangle diagram responsible for chiral non-conservation with one vertex of chiral current and two vertices of photons. On the right: The effect of interactions on chiral non-conservation can be schematically shown by a Gaussian distribution of the photon legs. }
    \label{fig:Triangle}
\end{figure}

So far we have exploited only the non-perturbative language, it is worthwhile however, to discuss the results in perturbative a language as well. Consider the vacuum expectation value of the chiral current $j^\mu_5$ by utilizing the path-integral \eqref{DecoupledI},
\begin{equation}
	\langle \bar{\psi} \gamma^\mu \gamma_5 \psi \rangle_{a;\psi} \equiv \int \mathcal{D}a \int \mathcal{D}\bar{\psi}\mathcal{D}\psi \, \bar{\psi} \gamma^\mu \gamma_5 \psi \, e^{S_a} \, ,
	\label{VevChiral}
\end{equation}
where the subscript $a$ designates that we have separated the integral over auxiliary field $a_\mu$ from the rest of the integrals denoted by the subscript $\psi$. We then notice that
\begin{equation}
	\langle \bar{\psi}(x) \gamma^\mu \gamma_5 \psi (x) \rangle_\psi = -\lim_{y \to x} \langle T^\star \gamma^\mu_{\alpha\sigma} \gamma_{5\sigma\beta} \psi_\beta (x) \bar{\psi}_\alpha (y) \rangle_\psi \, , \nonumber
\end{equation}
where $\alpha$, $\beta$ and $\sigma$ are running over spinor indices of the gamma matrix and fermionic operators while $T^\star$ is the time ordering operator. The two point function on the write hand side can moreover be written as,
\begin{align}
	\langle T^\star \psi (x) \bar{\psi} (y) \rangle_\psi &= \frac{1}{i\slashed{D}^x_g} \frac{1}{Z} \int \mathcal{D}[\bar{\psi}\psi] i\slashed{D}^x_g \psi(x)\bar{\psi}(y) e^{S_a} \nonumber \\
	&= \frac{1}{i\slashed{D}^x_g} \frac{1}{Z} \int \mathcal{D}[\bar{\psi}\psi] \frac{\delta \, e^{S_a}}{\delta \bar{\psi}(x)} \bar{\psi}(x) \nonumber \\
	&= - \frac{1}{i\slashed{D}^x_g} \frac{1}{Z} \int \mathcal{D}[\bar{\psi}\psi] \frac{\delta \bar{\psi}(y)}{\delta \bar{\psi}(x)} e^{S_a} = - \frac{1}{i\slashed{D}^x_g} \delta (x-y) \, .
\end{align}
Here $\slashed{D}^x_g$ is the generalized Dirac operator introduced previously containing a partial derivative that acts on functions of $x$, and $1/\slashed{D}^x_g$ just represent the inverse of that operator. Knowing this we can rewrite \eqref{VevChiral} as
\begin{equation}
	\langle \bar{\psi} \gamma^\mu \gamma_5 \psi \rangle_{a;\psi} = \lim_{y \to x} \langle \tr [ \gamma^\mu\gamma_5 (i\slashed{D}_g)^{-1} \delta(x-y) ] \rangle_a \, ,
	\label{<j5>}
\end{equation}
where the trace is over spinor indices. By remembering the definition of the generalized Dirac operator $i\slashed{D}_g = i\slashed{\partial} + e\slashed{A} + \lambda \slashed{a}$ we can expand $\slashed{D}_g^{-1}$ in powers of $e\slashed{A} + \lambda \slashed{a}$ and then employ the integration of the auxiliary field $a_\mu$ over each order. The expansion of $\slashed{D}_g^{-1}$ has the following form,
\begin{align}
	\frac{1}{i\slashed{D}_g} &= \frac{1}{i\slashed{\partial}} - \frac{1}{i\slashed{\partial}}  (e\slashed{A}+\lambda\slashed{a})\frac{1}{i\slashed{\partial}} + \frac{1}{i\slashed{\partial}}  (e\slashed{A}+\lambda\slashed{a}) \frac{1}{i\slashed{\partial}}  (e\slashed{A}+\lambda\slashed{a}) \frac{1}{i\slashed{\partial}} + \dots \, ,
	\label{PropagatorExpansion}
\end{align}
the last term of which gives rise to the triangle diagram, Fig. \ref{fig:Triangle}. In hindsight we know that only this term contributes to the chiral anomaly and all other higher order terms cancel out~\cite{AdlerBardeen}. The difference from the non-interacting case is the new element $\slashed{a}$ that appears because of the local current-current interactions alongside $\slashed{A}$. Without disturbing the triangle nature of chiral anomalies, this new term will nevertheless contribute to the chiral symmetry breaking. Integration on $a_\mu$ is Gaussian with a peak on $a_\mu = 0$. Therefore, the triangle diagram coming from \eqref{PropagatorExpansion} can be looked at as having a Gaussian distribution of photon legs peaked at $(a_\mu + A_\mu) = A_\mu$. (See Fig. \ref{fig:Triangle}.)

The chiral anomaly is given by substituting the last term of \eqref{PropagatorExpansion} in place of $(i\slashed{D}_g)^{-1}$ in \eqref{<j5>} and taking the divergence of the whole equality, or in momentum space, contracting it with an external momentum.

Instead of investigating the chiral anomaly by looking directly at the divergence of the chiral current, we can similarly investigate it through the response of a chiral system by considering the number current $j^\mu$. For this, we first introduce a constant axial field $\slashed{b}\gamma_5$ to the generalized Dirac operator and again expand $(i\slashed{D}_g)^{-1}$ in powers of $e\slashed{A} + \lambda\slashed{a}$ which to first order is given by,
\begin{equation}
	\frac{1}{i\slashed{D}_g} = \frac{1}{i\slashed{\partial}-\slashed{b}\gamma_5} - \frac{1}{i\slashed{\partial}-\slashed{b}\gamma_5}  (e\slashed{A}+\lambda\slashed{a})\frac{1}{i\slashed{\partial}-\slashed{b}\gamma_5} +\dots \, .
	\label{AxialPropagatorExpansion}
\end{equation}
Similar to \eqref{<j5>} we have a relation for $\langle j^\mu \rangle$ which after using the above expansion and successive Fourier transformations, will be as below
\begin{equation}
	\langle j^\mu \rangle_{a;\psi} = \left\langle \int_{k,q} \!\! \tr \left[\gamma^\mu \frac{-e^{-i q_\nu x^\nu}}{\slashed{k}+\slashed{q}-\slashed{b}\gamma_5}(e\slashed{A}+\lambda\slashed{a})\frac{1}{\slashed{k}-\slashed{b}\gamma_5} \right] \right\rangle_a \! ,
\end{equation}
where $\int_{k,q}$ is defined as $\int \frac{d^4 k}{(2\pi)^4} \frac{d^4 q}{(2\pi)^4}$ and the first term of the expansion \eqref{AxialPropagatorExpansion} has vanished. We can see, for example by multiplying the numerator and denominator of the last fraction by $\slashed{k}+\slashed{b}\gamma_5$, that additional linear in $A_\mu$ terms will appear due to the axial field $b_\mu$. These are of the following form
\begin{align}
	\langle j^\mu \rangle_{a;\psi} =  \bigg\langle \int_{k,q} f(k,q) \tr [ & e\gamma^\mu\gamma^\alpha\gamma^\nu\gamma^\beta \gamma_5 q_\alpha A_\nu(q) b_\beta \nonumber \\
	&+ \lambda\gamma^\mu\gamma^\alpha\gamma^\nu\gamma^\beta \gamma_5 q_\alpha a_\nu(q) b_\beta ]  \bigg\rangle_a +\dots \, .
\end{align}
with $f(k,q)$ being some function of the two four-momentums $k^\mu$ and $q^\mu$. Since for the Euclideanized picture we have $\tr[\gamma^\mu\gamma^\alpha\gamma^\nu\gamma^\beta\gamma_5]=4\epsilon^{\mu\alpha\nu\beta}$ the vacuum expectation value of the current density $j^\mu (x)$ acquires a contribution from $\epsilon^{\mu\alpha\nu\beta}b_\alpha \partial_\nu A_\beta$ and another from $ \epsilon^{\mu\alpha\nu\beta} b_\alpha \partial_\nu \langle j_\beta \rangle$. We will discuss the exact path-integral treatment of the above in section \ref{sec:Measurement} when we calculate the response as a way to measure the effect of interactions on chiral anomaly.

\section{Dimensional Reduction} \label{sec:DimRed}

The chiral symmetry breaking relation in presence of interactions i.e. equation \eqref{FullWard}, can be simplified when the system admits certain symmetries. For instance, when both electric and magnetic fields are kept parallel to the $z$-axis, $\vec{E}=E_z \hat{z}$ and $\vec{B}=B_z \hat{z}$, we will have rotational and translational symmetry across $x$--$y$ plane. This insures us that on average the currents and derivatives along $x$ and $y$ directions vanish. Therefore the last term in equation \eqref{FullWard} vanishes and it reduces to
\begin{equation}
	\partial_\mu j^\mu_5=\frac{e^2}{16\pi^2}\epsilon^{\mu\nu\rho\sigma} F_{\mu \nu}F_{\rho \sigma}-\frac{e B_z}{2\pi^2}\lambda^2_{\sigma\alpha}\epsilon^{12\rho\sigma}  \partial_\rho j^\alpha \, .
\end{equation}

The magnetic field along $\hat{z}$ generates Landau levels in a non-interacting theory. It would have been surprising if raising interaction strengths from zero to a small value would destroy the Landau level structure completely. Thus it is quite reasonable to assume that for small interaction strengths or large magnetic fields there still exist Landau levels and therefore also a lowest Landau level (LLL). The contribution of all Landau levels to chiral anomaly are canceled out due to the spin degeneracy of each level, except for the lowest Landau level which does not suffer from such degeneracy. As we have discussed before (for example around equation \eqref{RawConservation} and its following paragraph) the chiral anomaly comes from the zero-modes and the current situation is no exception.

Let us examine $j^\mu=\bar{\psi}\gamma^\mu\psi$ in this situation and in chiral representation: We can write,
\begin{align}
	j^\mu= \bar{\psi}\gamma^\mu\psi &=
		\begin{pmatrix}
			u^\dagger  & v^\dagger
		\end{pmatrix}
		\begin{pmatrix}
			\bar{\sigma^\mu} & 0 \\
			0 & \sigma^\mu
		\end{pmatrix}
		\begin{pmatrix}
			u \\
			v
		\end{pmatrix} = u^\dagger \bar{\sigma^\mu}  u + v^\dagger \sigma^\mu v \, ,
\end{align}
where $u$ and $v$ are two-component Weyl spinors that constitute Dirac spinors $\psi$, $\sigma^\mu\equiv (\mathbb{1},\sigma^k)$ and $\bar{\sigma}^\mu\equiv (\mathbb{1},-\sigma^k)$. When the spinors are situated at the LLL they can have definite spins along $\hat{z}$ leaving them with only one non-zero component in the basis of $\sigma^3$, and $j^\mu$ given as above, will have non-vanishing components only for $\mu=0$ and $3$. The same goes for the chiral current $j_5^\mu$. Considering all these we realize that the system has gone through a dimensional reduction from $(3+1)$ dimensions to $(1+1)$ dimensions. We can further see this by noticing that within zero-modes of our system, which is all we are concerned about here, the relation $\epsilon^{21\mu\nu}\gamma_\nu=\gamma^\mu\gamma_5$ holds if we keep $\mu \in \{0,3\}$ and $\nu \in \{0,3\}$. Compare this with the relation $\epsilon^{\mu\nu}\gamma_\nu=\gamma^\mu\gamma_5$ which only holds in $2$-dimensional spacetimes and was used to calculate equation \eqref{2DIntAnomaly}. After some rearranging we arrive at
\begin{equation}\label{reducedAnomaly}
    \partial_\mu j^\mu_5= \frac{1}{1+n_0\lambda^2_{3}/\pi}\frac{e^2}{2\pi^2}E_zB_z -\frac{n_0\left(\lambda^2_{0} - \lambda^2_{3}\right)/\pi}{1+n_0\lambda^2_{3}/\pi} \partial_3 j^3_5 \, ,
\end{equation}
where  $n_0 \equiv eB_z/2\pi$. Here we have also specialized to the case where the interaction tensor is diagonal.  In deriving this equation we have made no assumptions on the nature of Landau levels or how they arise, only that they exist which seems a physically reasonable proposition especially in the limit of large background field.   In the opposite limit of zero background field \eqref{reducedAnomaly} reduces to the noninteracting result.

Furthermore, when $\lambda_0^2=\lambda_3^2=\lambda^2$ e.g. for Lorentz symmetry respecting interactions, equation \eqref{reducedAnomaly} simplifies even more to
\begin{eqnarray}
	\!\!\!\!\!\!\!\!\!\!\! \partial_\mu j_5^\mu = \frac{1}{1+n_0 \lambda^2/\pi}\frac{e^2}{2\pi^2}E_z B_z = \frac{n_0}{1+n_0 \lambda^2/\pi}\frac{e}{\pi}E_z \, .
	\label{reducededAnomaly}
\end{eqnarray}
We see that here the chiral symmetry breaking in the absence of interactions has been modified by a coefficient much like \eqref{2DIntAnomaly} in $(1+1)$ dimensions. The only difference here is the quantity $n_0$ which is nothing but the degeneracy of the lowest Landau level per unit area.

The similarity of \eqref{reducededAnomaly} to the $(1+1)$-dimensional case \eqref{2DIntAnomaly} can be further expounded through describing a $(1+1)$-dimensional system with $N$ flavors of fermions in an all to all interacting relation. Consider the following path-integral,
\begin{align}
	I_N = \int &\mathcal{D}a_\mu \prod_{i=1}^N \mathcal{D}\bar{\psi}_i\mathcal{D}\psi_i \exp \left\{ \int d^2x \left[\sum_{i=1}^N \bar{\psi}_i i\slashed{D}_g \psi_i  +\frac{1}{2}a_\mu a^\mu \right] \right\} \, ,
	\label{PIofN}
\end{align}
with $\slashed{D} = \gamma^\mu\left(\partial_\mu - ieA_\mu - i\lambda a_\mu \right)$. We can separately define $N$ current densities $j_i^\mu \equiv \bar{\psi}_i \gamma^\mu \psi_i$ for each flavor. The on-shell value of $a_\mu$ is so given by $a^\mu = -\lambda \sum^N_{i=1} j^\mu_i$ which, when $a_\mu$ is integrated out, generates current-current interactions between all the current densities with coupling $-\lambda^2/2$ similar to what have done before.

Under the simultaneous chiral transformation of all flavors,
\begin{equation}
	\psi_i \longrightarrow e^{i\alpha\gamma_5} \psi_i, \ \ \ \ \  \bar{\psi}_i \longrightarrow \bar{\psi}_i e^{i\alpha\gamma_5} ,
	\label{ChiralTransN}
\end{equation}
the action functional of \eqref{PIofN} goes under the following change,
\begin{equation}
	\delta S_N = \int d^2x \, \alpha \partial_\mu \sum^N_{i=1} \bar{\psi}_i\gamma^\mu\gamma_5\psi_i \equiv \int d^2x \, \alpha\partial_\mu J_5^\mu \, ,
\end{equation}
hence $J_5^\mu$, defined as above, is classically conserved. However the measure, under each one of the chiral rotations in \eqref{ChiralTransN}, will introduce a copy of the $(1+1)$-dimensional chiral anomaly to break the classical conservation
\begin{equation*}
	\partial_\mu J_5^\mu = N \left( \frac{e}{\pi}E_z + \frac{\lambda}{\pi}\epsilon^{\mu\nu}\partial_\mu a_\nu \right) = N \frac{e}{\pi}E_z - N\frac{\lambda^2}{\pi}\partial_\mu J_5^\mu \, ,
\end{equation*}
where $e E_z/\pi$ is the anomalous term for one flavor in the absence of interactions and the last equality stems from integrating the auxiliary field $a_\mu$ out. For the $J_5^\mu$ in the last equality to appear we again have used the relation $\epsilon^{\mu\nu}\gamma_\nu = \gamma^\mu\gamma_5$ between $2$-dimensional gamma matrices. A simple rearranging of the above equation yields,
\begin{equation}
	\partial_\mu J_5^\mu = \frac{N}{\left( 1+N\lambda^2/\pi \right)} \frac{e}{\pi}E_z \, .
	\label{2DNAnomaly}
\end{equation}
Upon identifying $N$ with the degeneracy of LLL per area $n_0$, equations \eqref{2DNAnomaly} and \eqref{reducededAnomaly} become identical.

Moreover, we can take the same steps we took in section \ref{sec:EffAct} for these $N$ flavor fermionic system, to transform the action functional to
\begin{equation}
	S_N = \int d^2x \sum_{i=1}^N \bar{\Psi}_ii\gamma^\mu \left[ \partial_\mu -i\frac{e}{\sqrt{1+N\lambda^2/\pi}}A_\mu \right] \Psi_i \, ,
\end{equation}
which means that in $(1+1)$ dimensions we can treat an $N$-flavor interacting fermionic system as an $N$-flavor free fermionic system where the coupling to external electromagnetic field has been modified by $(1+N\lambda^2/\pi)^{-1/2}$. Consistently, if we start with a $(3+1)$-dimensional free fermionic system which has gone through a dimensional reduction due to the presence of parallel magnetic and electric fields, but has an electric charge given by
\begin{equation}
 \tilde{e} \equiv \frac{e}{\sqrt{1+\frac{eB_z}{2\pi}\frac{\lambda^2}{\pi}}} \, ,
\end{equation}
we get the same result as \eqref{reducededAnomaly}.

\section{Measurable Consequences} \label{sec:Measurement}

The models that we have been considering so far can be realized in many physical systems. In lattice systems of these sort the band structure forms Dirac cones where, around the point that the two low energy bands have deformed to touch each other, the description of the system is given by fermionic path-integrals such as $\int \mathcal{D}[\bar{\psi}\psi] \exp\{ i \int \! d^n \!\!\; x \, \bar{\psi}i\slashed{D}\psi \} $. One can point to graphene, topological insulators and liquid $^3\text{He}$ as few examples~\cite{Graphene, EEScales,TopolInsulator,Topol3D,TopolBirth,NonAbelianAnyons,VolovikExotic,VolovikUniverse}. When the energy dispersion has genuine doubly degenerate Dirac cones, they can be made separated into two chirally distinct cones by introducing a chiral or time-reversal breaking element. Among material that posses such feature, are Weyl semimetals~\cite{BalentsBurkov,TopologicalMaterialsWeyl,Zhou,GuoFan,Wang}---a type of gapless topological matter with distinctive features including a large negative magnetoresistance~\cite{NielsenNinomiya, SonSpivak,Burkov, FukushimaKharzeevWarringa} and an anomalous Hall response~\cite{ZyuzinBurkov,ChenWuBurkov}. The chiral element in Weyl semimetals, which separates the otherwise degenerate cones, appears in the low energy description as the additional term $\int d^4 x b_\mu j^\mu_5$ in the action functional with $b_\mu$ being constant. When projected to left and right handed spinors this term breaks into $\int d^4 x b_\mu (j^\mu_R - j^\mu_L)$ which clearly destroys the preexisting symmetry $L\leftrightarrow R$ between the exchange of left and right moving fermions.

\subsection{Prior to Interactions}

Before jumping to the interacting case, let us first derive the anomalous transport of a chiral system. The low energy description of the Weyl semimetal is provided by the following path-integral,
\begin{align}
	  Z_b[A_\mu]= \int \mathcal{D}[\bar{\psi}\psi] \exp\left\{\int d^4x \left[ \bar{\psi} i\slashed{D} \psi + b_\mu j^\mu_5 \right] \right\} \, ,
\end{align}
where as before $\slashed{D}=(\partial_\mu -ieA_\mu)$ with $A_\mu$ being an external field and spinor degrees of freedom are integrated over. Also the subscript in the partition function $Z_b$ indicates that it is carrying a chiral element (or Weyl separation) which breaks the $L\leftrightarrow R$ symmetry.

Since $b_\mu$ is constant we can always perform a chiral transformation to remove $-ib_\mu \gamma_5$ from inside the parentheses. This transformation is given by $\psi \rightarrow e^{ib_\mu x^\mu\gamma_5}\psi$, $\bar{\psi} \rightarrow \bar{\psi} e^{ib_\mu x^\mu\gamma_5}$. But of course the measure is not invariant under such transformation and introduces an additional term to the action,
\begin{equation}
	Z_b =\int \mathcal{D}[\bar{\psi}\psi] \exp\left\{\int d^4x \bigg [ \bar{\psi} i\slashed{D} \psi  - \frac{e^2}{4\pi^2} \epsilon^{\mu\nu\rho\sigma} b_\mu A_\nu  \partial_\rho A_\sigma  \bigg ] \right\} \, ,
\end{equation}
where the same path-integral, $Z_b[A_\mu]$, now comes with a different action in which the $b_\mu j^\mu_5$ has turned to a Chern-Simons like term. Here we can establish a relation between partition functions that carry the chiral element and the ones that do not:
\begin{align}
	&Z_b[A_\mu] = Z[A_\mu] \exp{\left\{- \frac{e^2}{4\pi^2}\int d^4x \,   \epsilon^{\mu\nu\rho\sigma} b_\mu A_\nu  \partial_\rho A_\sigma \right\} } \, , \nonumber \\
	& \mbox{with} \quad \quad \  Z[A_\mu] \equiv Z_b[A_\mu] \Big|_{b_\mu=0} \, .
	\label{ZZRelation}
\end{align}

We wish to know how much the chiral element contributes to the electrical current. To obtain the current in non-zero $b_\mu$ we vary $Z_b$ with respect to the external gauge field and write $ e \langle j^\mu \rangle_b = \frac{1}{Z_b}\frac{\delta Z_b}{\delta A_\mu}$. Correspondingly we have the current $e \langle j^\mu \rangle = \frac{1}{Z}\frac{\delta Z}{\delta A_\mu}$ when $b_\mu$ has been excluded. The response due to the chiral element, which we may call the anomalous response $j_A^\mu$, is obtained from the difference of these two. Varying the above relation between $Z_b$ and $Z$ with respect to the external gauge field $A_\mu$ gives us just that,
\begin{equation}
	e j_A^\mu = e \langle j^\mu \rangle_b- e \langle j^\mu \rangle = \frac{e^2}{2\pi^2}\epsilon^{\mu\nu\rho\sigma}b_\nu \partial_\rho A_\sigma \, .
	\label{AnomalousCurrent}
\end{equation}
Note that the anomalous current is conserved i.e. $\partial_\mu j^\mu_A = \frac{e}{2\pi^2}b_\nu \epsilon^{\mu\nu\rho\sigma}\partial_\mu\partial_\rho A_\sigma=0$.

\subsection{Interactions Included}

Having the current formula \eqref{AnomalousCurrent} at hand and remembering our experience in deriving equation \eqref{tildeF}, we can readily guess that substituting $eA_\sigma$ by $e\tilde{A}_\sigma = eA_\sigma - \lambda^2_{\sigma\alpha} \langle j^\alpha \rangle_b$ should give us the anomalous current in presence of interactions. Nevertheless, we are going to take more careful steps and calculate the interacting formula rigorously 
as follows.

The low energy description of an interacting Weyl material can then be given by the following path-integral
\begin{align}
	I_b =\int \mathcal{D}[\bar{\psi}\psi]\exp \left\{ \int d^4 x \left[ \bar{\psi} i\slashed{D} \psi +b_\mu j^\mu_5 -\frac{1}{2}\lambda^2_{\mu\nu} j^\mu j^\nu \right] \right\} \, , 
	\label{Ic}
\end{align}
As we have done so many times now, we are going to decouple the current-current interaction by introducing an auxiliary field $a_\mu$ to reform the path-integral as below,
\begin{align}
	I_b =\int &\mathcal{D}[\bar{\psi}\psi a_\mu] \exp \left\{ \int d^4x \left[ \bar{\psi} i \slashed{D}_g \psi +b_\mu j^\mu_5 -\frac{1}{2}a_\mu a^\mu \right] \right\} \, , 
\end{align}
where as before the generalized Dirac operator is given by $\slashed{D}_g \equiv \left( \partial_\mu -ieA_\mu  - i \lambda_{\mu\nu} a^\nu  \right)$ and the previous form of the path-integral is obtained by integrating over $a_\mu$.

We are going to exploit the fact that both $a_\mu$ and $b_\mu$ can decouple from fermions, the former by a shift with its on-shell value and the latter by a chiral rotation. Let us first decouple $b_\mu$ from fermions via the chiral rotation $\psi \rightarrow e^{ib_\mu x^\mu\gamma_5}\Psi$, $\bar{\psi} \rightarrow \bar{\Psi} e^{ib_\mu x^\mu\gamma_5}$.
All the terms in the action are invariant under this rotation, but the measure transforms with a non-trivial Jacobian and the path-integral becomes reorganized as
\begin{align}
	I_b =\int \mathcal{D}[\bar{\Psi}\Psi a_\mu] \exp \int d^4x \bigg [ &\bar{\Psi} i \slashed{D}_g \Psi  -\frac{1}{2}a_\mu a^\mu  \\
	&  - \frac{1}{4\pi^2} \epsilon^{\mu\nu\rho\sigma} b_\mu \left(eA_\nu + \lambda_{\nu\alpha}a^\alpha \right) \partial_\rho \left(eA_\sigma + \lambda_{\sigma\beta}a^\beta \right) \bigg ] \, . \nonumber
\end{align}
The relation corresponding to \eqref{ZZRelation} is given for the interacting case by
\begin{align}
	&I_b \equiv \int \mathcal{D}a_\mu Z_b [A_\mu,a_\mu] = \int \mathcal{D}a_\mu Z[A_\mu,a_\mu] \exp \left\{ -\frac{e^2}{4\pi^2}\int d^4x \,   \epsilon^{\mu\nu\rho\sigma} b_\mu \mathcal{A}_\nu  \partial_\rho \mathcal{A}_\sigma \right\} \, , \nonumber \\
	&\mbox{with} \quad \quad Z_b[A_\mu,a_\mu]= \int \mathcal{D}[\bar{\psi}\psi] \exp\left\{\int d^4x \left[ \bar{\psi} i\slashed{D}_g \psi + b_\mu j^\mu_5 - \frac{1}{2}a_\mu a^\mu\right] \right\} \, ,  \nonumber \\
        &\text{and } \quad \quad Z [A_\mu,a_\mu]\equiv Z_b [A_\mu,a_\mu]\Big|_{b_\mu=0} \, , \label{ZZRelationI}
\end{align}
where $e \mathcal{A}_\mu \equiv e A_\mu + \lambda_{\mu\alpha} a^\alpha$. Therefore, similar to \eqref{AnomalousCurrent}, upon varying \eqref{ZZRelationI} with respect to $A_\mu$, we have the following relation for currents,
\begin{equation}
	\int \mathcal{D}a_\mu \langle j^\mu \rangle_b \, Z_b [A_\mu,a_\mu] = \int \mathcal{D}a_\mu \left( \langle j^\mu \rangle + \frac{e}{2\pi^2}\epsilon^{\mu\nu\rho\sigma}b_\nu \partial_\rho \mathcal{A}_\sigma \right) \, Z_b [A_\mu,a_\mu] \, .
\end{equation}
We now rotate the fermionic degrees of freedom $\Psi$ back to their original state $\psi$ by the reverse chiral rotation. Doing this leaves $Z_b$ and $j^\mu$ and consequently everything in the above relation unchanged, but is nevertheless crucial, first because the relevant correlation functions are those expressed in terms of the original fermions and second for making integration over the auxiliary field $a_\mu$ possible. 

Now it is time to decouple $a_\mu$ from fermions by the shift $a_\mu \rightarrow a_\alpha - \lambda_{\beta\alpha}j^\beta$ which results in getting the current-current interaction in \eqref{Ic} back. Note that the reverse chiral rotation to original fermions has removed the Chern-Simons like term from the action which now after the shift in the auxiliary field is quadratic in $a_\mu$. This means that the term $\epsilon^{\mu\nu\rho\sigma}b_\nu\partial_\rho (\lambda_{\sigma\alpha}a^\sigma)$, which is linear in $a_\mu$, will be eliminated from $\epsilon^{\mu\nu\rho\sigma}b_\nu \partial_\rho \mathcal{A}_\sigma$ after integration over the auxiliary field. At this stage both current expectation values and $Z_b$ are untied from $a_\mu$ meaning that $a_\mu$-terms can be factored out of them. Thus we are left with
\begin{equation}
 ej_A^\mu = \frac{e}{2\pi^2}\epsilon^{\mu\nu\rho\sigma}b_\nu\partial_\rho\left(e A_\sigma - \lambda^2_{\sigma\alpha} \langle j^\alpha \rangle_b \right) \, .
 \label{AnomalousCurrentI}
\end{equation}
This result is therefore true for all $\lambda_{\mu\nu}$.

\subsection{Anomalous Transport}

Remember that the total current, $\langle j^\mu \rangle_b$, is the sum total of the anomalous current, $j_A^\mu$, and the non-anomalous current. When the non-anomalous part of the total current is vanishing, we can summarize the result \eqref{AnomalousCurrentI} as $j^\mu = \frac{e^2}{4\pi^2}\epsilon^{\mu\nu\rho\sigma}b_\nu \tilde{F}_{\rho\sigma}$ with $\tilde{F}_{\mu\nu}$ defined under equation \eqref{tildeF}. Let us simplify the current equation by specializing to the case of $b^\mu=b_z \delta^\mu_3$ and Lorentz invariant interaction $\lambda_{\mu\nu}=\lambda\eta_{\mu\nu}$, which gives us a $2b_z$ separation of Weyl nodes along $k_z$ direction in momentum space and natural current-current interaction which is but the density-density interaction made Lorentz invariant. When only $b_3$ is non-zero, $j^3 \equiv j^z$ is zero and we have for other spatial directions $ej^x = \sigma^{xy}_0 \tilde{E}^y$ with $\sigma^{xy}_0=e^2 b_z/2\pi^2$ and similarly for $j^y$. As we can see from the definition \eqref{D}, at the equilibrium and homogeneous limit $\tilde{\vec{E}}$ goes to $\vec{E}$. Therefore at such limit, we end up with the same anomalous Hall response formula as in the non-interacting case. More explicitly,
\begin{equation}\label{HallResponse}
  j^x=\frac{eb_z}{2\pi^2}E^y-\lambdabar\left[\partial_t j_y-\partial_{y}\rho\right] \, ,
\end{equation}
with $\lambdabar\equiv \lambda^2 b_z / 2\pi^2$. The first term gives the quantum anomalous Hall current in either non-interacting materials or the equilibrium and homogeneous limit; in both cases the second term vanishes. Therefore, even though interactions do not affect the equilibrium Hall current, they may nevertheless contribute to the nonequilibrium or inhomogeneous response.

To obtain the Hall conductivity in the homogeneous but nonequilibrium limit we combine equation \eqref{HallResponse} with the corresponding expression for $j^y$ and after switching to Fourier space we get
\begin{equation}
	\sigma^{xy}(\omega)=\left[1+\left(\lambdabar \omega\right)^2\right]^{-1}\sigma^{xy}_0 \, ,
\end{equation}
where again, $\sigma^{xy}_0 = e^2b_z/2\pi^2$ is the Hall conductivity in the absence of interactions. There is also a contribution to the longitudinal conductivity arising solely due to interplay of interactions with the Hall conductivity,
\begin{equation} \label{LongResponse}
	\! \sigma^{xx}(\omega)= \frac{\lambdabar \omega}{1+\left(\lambdabar \omega\right)^2}  \sigma^{xy}_0 =  \lambdabar \omega  \sigma^{xy}(\omega) \, .
\end{equation}
For small $\omega$ i.e small deviations from equilibrium, the leading order in $\sigma^{xx}(\omega)$ is simply $\lambdabar \omega \sigma_0^{xy}$, whereas at large enough $\omega$, given the current formulation holds, the anomalous longitudinal conductivity $\sigma^{xx}(\omega)$ vanishes along with $\sigma^{xy}(\omega)$. It is noteworthy that had the non-anomalous part of the current been non-zero its contribution could appear as an additional non-anomalous longitudinal conductivity $\sigma_0^{xx}$ on the right hand side of \eqref{LongResponse}.\footnote{We can see this simply by substituting $j_A^\mu$ with $\langle j^\mu \rangle_b - \langle j^\mu \rangle$ in equation \eqref{AnomalousCurrentI} and then writing the spatial part of the non-anomalous current as conductivity times electric field, in particular $\langle j^x \rangle = \sigma_0^{xx} E^x$. Then it becomes clear that whatever the anomalous or Hall conductivity is, the total conductivity is obtained by adding $\sigma^{xx}_0$ to its longitudinal part.}

But the anomalous Hall effect is not the only anomalous transport phenomena that might be influenced by the effects of interactions. What about other responses at other limits, do, for example, interactions affect the equilibrium density response to a change of the magnetic field? To see if that is the case, it is both convenient and insightful to look back at the dimensionally reduced system, section \ref{sec:DimRed}, where as we have calculated in \eqref{reducededAnomaly} the effect of interactions on chiral symmetry breaking is simply a factor that depends on the degeneracy $n_0 = eB_z/2\pi$ of LLL while a strong magnetic field has already generated a background charge density. We start from the path-integral \eqref{Ic} in the dimensionally reduced setup where the full Lorentz symmetry is broken into a rotational symmetry in $x$-$y$ plane and a boost symmetry along $t$-$z$, with $\lambda^2_{\mu\nu}=\lambda^2(\eta_{0\mu}\eta_{0\nu} + \eta_{3\mu}\eta_{3\nu})$, $\vec{E}=E_z \hat{z}$ and $\vec{B}=B_z \hat{z}$. We also restrict $b_\mu$ to $b_z\delta_\mu^3$ as before. Removing $b_z j^z_5$ from the Lagrangian by the chiral rotation employed in previous section, adds to the action the right hand side of the following relation.
\begin{align}
	\int d^4x \, b_z j^z_5 =
	 \frac{e^2/4\pi^2}{1+n_0 \lambda^2/\pi} \int d^4x \, \epsilon^{\nu 3 \rho\sigma} b_z A_\nu\partial_\rho A_\sigma \, .
	 \label{reducedCS}
\end{align}
The above relation can be confirmed by integrating over the modified chiral charge conservation law \eqref{reducededAnomaly} multiplied by the parameter of the chiral rotation $b_\mu x^\mu$. Now that we have the right hand side of \eqref{reducedCS} substituting $\int d^4 x b_z j^z$ inside the action, varying the path-integral with respect to $A_0 (x)$ will give us the anomalous density:
\begin{equation}
	e j^0_A=e \rho_A=\frac{n_0}{1+n_0\lambda^2/\pi}\frac{e}{\pi}b_z \, .
\end{equation}
Remember that $n_0$ depends on $B_z$. Also note that  in reaching the above equation we have made no assumptions regarding the non-anomalous part of the total current.

If we change the background magnetic field by a small amount $B_z \rightarrow B_z + \delta B_z$ and then let the system go to the new equilibrium, the background density will be altered by a small amount. For the case where $\lambda^2$ is positive, it is given to first order by
\begin{equation}
	e\delta\rho_A=\frac{\delta n_0}{\left(1+n_0 \lambda^2/\pi\right)^2}\frac{e}{\pi}b_z \, ,
	\label{DensityResponse}
\end{equation}
with $\delta n_0 = e\delta B_z/2\pi$. Here, if we keep the magnetic field large enough so that LLL is formed, there are two domains of magnetic field strength that the density transport phenomena has distinct behaviors in. For large $B_z$ limit there is no density response to a change of magnetic field, whereas at low enough $B_z$ the density responds linearly.

Notice that due to dimensional reduction the chiral current in the longitudinal direction $\langle j^z_5 \rangle$ is equivalent to density $\langle \rho \rangle$ through the relation $\epsilon^{\mu\nu}\gamma_\nu = \gamma^\mu \gamma_5$ which holds in two dimensional spacetimes. Therefore, equation \eqref{DensityResponse} can be viewed as the generation of a chiral current in response to a change in the magnetic field which is known as the chiral separation effect (CSE)~\cite{Vilenkin, MetlitskiZhitnitsky, NewmanSon}.

At this point we briefly mention that the current formula \eqref{AnomalousCurrentI} can be also exploit for the interacting chiral magnetic effect when separation of Weyl point has a temporal component $b_t \neq 0$. Moreover, investigations through bosonization and random phase approximation shed more light on the subject. These are discussed in reference~\cite{CAICMS}.

\section{Anomalous Modes} \label{sec:DynamicAnomaly}

In this section we briefly turn our attention to the interacting anomalous current formula \eqref{AnomalousCurrentI} to better uncover the dynamics that resides inside it. Turning the external electromagnetic field off makes it clear that the interplay between interactions and anomaly alone is creating a dynamical behavior among anomaly generated phenomena. Let us then set $A_\mu =0$, and for simplicity choose $b_\mu = b_z \delta^3_\mu$ in addition to the Lorentz invariant interaction $\lambda^2_{\mu\nu}=\lambda^2\eta_{\mu\nu}$, to obtain out of equation \eqref{AnomalousCurrentI} the following formula,
\begin{equation}
	j^\mu = \lambdabar \epsilon^{3\mu\alpha\beta} \partial_\alpha j_\beta \, ,
	\label{DynamicCurrent}
\end{equation}
where we have dropped the subscript in $j^\mu_A$ since we are only considering the situation where only the anomalous current is present, and again we are using $\lambdabar \equiv \lambda^2 b_z / 2\pi^2$ notation as in the previous section.

Since in $b_\mu$ only the $\mu=3$ component is non-zero the Levi-Civita tensor makes sure that $j^z$ is zero. The other three equations are
\begin{align}
	j^x &= \lambdabar \left( \partial_y \rho - \partial_t j_y \right) \label{jx} \\
	j^y &= \lambdabar \left( \partial_t j_x -\partial_x \rho \right) \label{jy} \\
	\rho &= \lambdabar \left( \partial_x j_y - \partial_y j_x \right) = \lambdabar \vect{\nabla} \times \vect{j} \label{rho} \, .
\end{align}
The first two equations are the ones that give rise to dynamical behavior, while the third one is a constraint equation involving no time-derivative. Here the curl operator is in nature a two-dimensional differential operator in $x$-$y$ plane. But since $j^z=0$ is kept zero and there are no equations, hence no dynamics, along $z$-direction we can pretend that differential operators are three-dimensional whenever it is desirable.

To see how charge disappears from a region let us take the time derivative of \eqref{rho} and write $\partial_t \rho = -\lambdabar \partial_t \vect{\nabla} \times \vec{j}$. On the other hand, by adding up the partial derivatives of \eqref{jx} and \eqref{jy} respectively along $x$ and $y$, we have $\vect\nabla\cdot\vect j = \lambdabar \partial_t \vect\nabla \times \vect j$. Comparing these two relations tells us that the anomalous charge is locally conserved:~$\partial_t \rho = - \vect\nabla \cdot \vect j$. This of course should come as no surprise, since we are reiterating the fact that $\partial_\mu j^\mu$ vanishes due to commutativity of partial derivatives and anti-commutativity of the indices of the Levi-Civita tensor in \eqref{AnomalousCurrentI} or \eqref{DynamicCurrent}.

Furthermore, incorporating dynamical equations \eqref{jx} and \eqref{jy} in the constraint equation \eqref{rho} gives $\rho =  \lambdabar^2 \left(  \partial_t \vect\nabla \cdot \vect j + \vect\nabla^2 \rho \right)$. Using the conservation of charge this becomes an equation for the density:~$\rho = \lambdabar^2 \left(   \vect\nabla^2 \rho - \partial_t^2 \rho \right)$. The similar relation holds for $j^x$ and $j^y$. Thus we end up with three Klein-Gordon equations for each non-zero component of $j^\mu$,
\begin{equation}
	\left( \partial_t ^2 - \vect\nabla^2 + \lambdabar^{-2} \right) j^\mu =0 \, .
	\label{KGwoA}
\end{equation}
Components of $j^\mu$ only depend on each other through the continuity equation, otherwise they have separate dynamics. Solutions to the above equation are relativistic propagating distributions $j^\mu = f^\mu(\omega t - \vec{ k\cdot  x})$ where $\omega$ and $k$ satisfy the relation $\omega^2 - k^2 = \lambdabar^{-2}$. Here $\lambdabar$ is the reduced Compton wavelength of these waves. By allowing the Fermi velocity $v_F$ and $\hbar$ to reappear, we can read off from $\lambdabar = \hbar/m v_F$ the mass attributed to these relativistic waves. In terms of the parameters of our model then,
\begin{equation}
    m=\hbar/\lambdabar v_F =h^2/2v_F \lambda^2 b_z \, . 
\end{equation}
So at the limit of very strong interactions or well separated Weyl points, the waves become massless and propagate with Fermi velocity.

Electromagnetic fields, then, are sources to equation \eqref{KGwoA} and we are going to confirm this by using equation \eqref{AnomalousCurrentI} on itself. Let us for the sake of accessibility rewrite it below
\begin{equation}
	j^\mu = \epsilon^{\mu\rho\sigma 3} \left(\frac{eb_z}{2\pi^2} \partial_\rho A_\sigma - \lambdabar \partial_\rho j_\sigma \right) \, .
\end{equation}
We are going to restrict all indices to $\{0,1,2\}$ since there is no need to bother about the third component. Using above equation twice gives,
\begin{align}
	j^\mu &= \epsilon^{\mu\rho\sigma 3}  \frac{eb_z}{2\pi^2} \partial_\rho A_\sigma - \lambdabar\left(\eta^{\mu\alpha}\eta^{\rho\beta} - \eta^{\mu\beta}\eta^{\rho\alpha}\right) \left(\frac{eb_z}{2\pi^2} \partial_\alpha A_\beta - \lambdabar \partial_\alpha j_\beta \right) \, .
\end{align}
By employing the conservation of current and after some rearranging we have,
\begin{equation}
	\left(1 + \lambdabar^2 \Box \right) j^\mu =  \frac{eb_z}{4\pi^2} \left( \epsilon^{\mu\rho\sigma 3} F_{\rho\sigma} +  2\lambdabar \partial_\rho F^{\rho\mu} \right) \, ,
	\label{Mawell}
\end{equation}
where we recall $\Box \equiv \partial_t^2 - \vect\nabla^2$ is the d'Alembertian. Again, the above equation describes three massive waves sourced by the electromagnetic field.

As an aside, note that the role of speed of light is played by the Fermi velocity $v_F \neq c$ in the d'Alembertian. Was it not the case, shining electromagnetic waves on the material would have a rather similar effect as in the non-interacting case. Let us for a moment set $v_F=c$ and then apply the inverse of $(1+\lambdabar^2\Box)$ from the left on both sides of the above equation. In effect, doing this removes the the d'Alembertian, since $\Box F_{\mu\nu}$ vanishes in this case.

The origin of the gauge field $A_\mu$ was external, meaning that no dynamical terms were introduced for the gauge field inside the action and it was treated as a source without being integrated over. Now it is rather interesting that regardless of the way the electromagnetic field started, Maxwell's equations have forced their way through, utilizing the interplay of chiral anomaly and interactions. Notice the last term in \eqref{Mawell} vanishes either when interactions are absent or no chiral element $b_\mu$ exists, while the second to last term is a purely axionic one. We can therefore conclude that there exist massive modes coupled to electromagnetic fields obeying axionic electrodynamics all due to the interaction-anomaly interplay.




\section{Introducing Gravity} \label{sec:Gravity}

Using what we have so far constructed, it is easy to see that a mixed chiral-gravitational anomalous relation in (3+1)-dimensions~\cite{Witten} will also be modified in a similar way. In the presence of curvature but absence of interactions, the four dimensional anomalous relation will host an additional geometrical Pontryagin density,
\begin{equation}
	\nabla_\mu j_5^\mu = \frac{e^2}{16\pi^2}\epsilon^{\mu\nu\rho\sigma}F_{\mu\nu}F_{\rho\sigma} + \frac{1}{384\pi^2}\epsilon^{\mu\nu\rho\sigma}R_{\mu\nu}^{\ \ \ \alpha\beta} R_{\rho\sigma\alpha\beta} \, ,
	\label{WardCurved}
\end{equation}
with $R^\mu_{\ \nu\alpha\beta}$ being the Riemann curvature tensor. We need to be more careful about curved space notions such as the covariant derivative $\nabla_\mu$ and the fact that $\epsilon^{\mu\nu\rho\sigma}$ is a Levi-Civita \textit{tensor} defined as $\varepsilon^{\mu\nu\rho\sigma}/\sqrt{|g|}$ with $\varepsilon^{\mu\nu\rho\sigma}$ being the totally anit-symmetric Levi-Civita \textit{symbol} and $g$ the determinant of the curved spacetime metric $g_{\mu\nu}$.\footnote{The covariant derivative $\nabla_\mu \equiv \partial_\mu + \Gamma_\mu$ is defined with respect to the object it is acting on. For spinor $\Gamma_\mu$ carries spinor indices while for vector or higher rank tensors it only has spacetime components. The Riemann tensor is then defined as,
\begin{equation}
    R^\mu_{\ \nu \alpha  \beta} A_\mu = -\left[\nabla_\alpha,\nabla_\beta\right] A_\nu \, .
\end{equation}
The Levi-Civita \textit{symbol}, $\varepsilon^{\mu\nu\alpha\beta}$ is totally anti-symmetric with $\varepsilon^{0123}=1$ and all other components given by permutations and the anti-symmetry constraint. This \textit{symbol} is a tensor density and it can be made into a tensor if it is divided by the square root of the metric determinant. 
}
These of course do not change our previous results in flat spacetime. This can be verified by considering the fact that we can always transform the coordinates to a frame where metric looks locally flat at any given point on spacetime.

The reason the gravitational term appears in the non-conservation of the chiral current is clear by the arguments made in subsection \ref{sec:Regularization}: When there is curvature the covariant derivative must accommodate its presence, leading to a generalized Dirac operator which also contains a spin-connection, $\slashed{D}_g \equiv \gamma^\mu (\partial_\mu - ieA_\mu - i\Gamma_\mu )$, with $\Gamma_\mu$ representing the spin-connection and $\gamma^\mu(x) = e^\mu_k(x) \gamma^k$ being the curved version of the flat gamma matrices $\gamma^k$. Also $e^\mu_k$ are the vielbeins satisfying $g^{\mu\nu}=e^\mu_m e^\nu_n \eta^{mn}$. The basis of the generalized Dirac operator formally diagonalizes the action and can be used to regularize the fermionic path-integral as well as the Jacobian of fermion transformations \eqref{RegA5}. In this way the curvature effects appear in the anomalous term and we arrive at equation \eqref{WardCurved}. (To better clarify the notation used above it should be added that we use Latin indices for objects that belong to the flat tangent space. Also, $\Gamma_\mu$ is a compact notation which encapsulates four matrices defined as $\Gamma_\mu \equiv \Gamma^{mn}_\mu [\gamma_m,\gamma_n]/4$ with the commutator of gamma matrices carrying all the spinor indices.)

We can treat the presence of interactions as before by a Hubbard-Stratonovich decoupling. Let us consider the same four-fermionic interaction term $\lambda^2_{\mu\nu}j^\mu j^\nu$ as in section \ref{sec:4D}. When curvature effects are included this leads to the following anomalous relation,
\begin{align}\label{FullCurvedWard} \nonumber
	\nabla_\mu j^\mu_5 =&\frac{\epsilon^{\mu\nu\rho\sigma}}{4\pi^2}\left( \lambda^2_{\nu\alpha}\lambda^2_{\sigma\beta}\nabla_\mu j^\alpha \nabla_\rho j^\beta - 2e\lambda^2_{\sigma\alpha}\nabla_\mu A_\nu\nabla_\rho j^\alpha \right) \\
	&  + \frac{e^2}{16\pi^2} \epsilon^{\mu\nu\rho\sigma} F_{\mu \nu}F_{\rho \sigma} + \frac{1}{384\pi^2}\epsilon^{\mu\nu\rho\sigma}R_{\mu\nu}^{\ \ \ \alpha\beta} R_{\rho\sigma\alpha\beta}  .
\end{align}
As is apparent these types of interactions do not produce cross-terms with curvature.

Perhaps the effects of interactions on the chiral-gravitational part of the anomalous term are most easily seen after a dimensional reduction as in section \ref{sec:DimRed} where the electric and magnetic fields are both pointing along $\hat{z}$ direction. Doing so and choosing $\lambda^2_{\mu\nu} = \lambda^2 \eta_{\mu\nu} $ greatly simplifies the above relation to,
\begin{align} \label{WardCurvedInt}
	&\left( 1 + n_0\frac{\lambda^2}{\pi} \right)  \nabla_\mu j_5^\mu = \frac{e^2}{16\pi^2}\epsilon^{\mu\nu\rho\sigma}F_{\mu\nu}F_{\rho\sigma} + \frac{1}{384\pi^2}\epsilon^{\mu\nu\rho\sigma}R_{\mu\nu}^{\ \ \ \alpha\beta} R_{\rho\sigma\alpha\beta} \, . 
\end{align}
When we are interested in the non-conservation of the chiral current $\nabla_\mu j^\mu_5 = \mathcal{A}_5$ we see that it has been modified by a factor of $(1+n_0 \lambda^2/\pi)^{-1}$ where we recall that $n_0$ is defined as $eB_z/2\pi$ in section \ref{sec:DimRed}.

\subsection{Gravity and Temperature}

Now that we have obtained the interacting anomalous relation in presence of curvature \eqref{FullCurvedWard} and \eqref{WardCurvedInt} we can discuss the corresponding response and possible measurable consequences which are the subject of the next three subsections. First we are going to consider two examples that provide an intuition about the geometrical part of the anomalous term.

For the winding density $R\tilde R \equiv \varepsilon^{\mu\nu\rho\sigma} R_{\mu\nu}^{\ \ \ \alpha\beta}R_{\rho\sigma\alpha\beta}$ to be non-zero, there should exist some sort of ``twist'' in the geometry. A spherical symmetric geometry, for instance, has a vanishing $R\tilde R$. Consider the following line element,
\begin{equation}
    ds^2 =  -dt^2 + dr^2 + dz^2 +  r^2 d\phi^2 - 2 r\Omega \Theta (z \! + \! l,l \! - \! z) dt d\phi \, .
\end{equation}
with $\Theta (z+l,l-z)$ being a generalized Heaviside step function: Zero whenever any of its arguments are negative and equal to identity otherwise. A non-vanishing $\Omega$ generates a difference between an angular step forward $d\phi>0$ and backward $d\phi<0$. So the line element describes a twist withing the region $-l \leq z \leq l$. For this metric we have $R\tilde R = 2[\delta (l+z) - \delta (l-z) ]\times \Omega^3/r^2(1+\Omega^2)^2$ which is non-zero only for non-vanishing $\Omega$. For a general $z$
 dependent $\Omega (z)$ we would have $R\tilde R = 8\Omega^2 \partial_z \Omega / r^2(1+\Omega^2)^2$ instead.
 
 As the second example consider the following metric,
\begin{equation}
    ds^2 =  -dt^2 + dr^2 + \left( dz - \Omega(t) d\phi \right)^2 + r^2 d\phi^2 \, .
\end{equation}
Let us define $d\tilde z \equiv dz-\Omega d\phi$, where now $\Omega(t)$ is an arbitrary function of time, and take $\tilde z$ as the new substitute for the $z$ axis which sets up a coordinate system where metric becomes diagonalized. Consequently, a step in $\phi$ while other coordinates $(t,r,\tilde z)$ are kept fixed, would mean a step in $z$ direction. This describes a spiral along $\hat z$. For this geometry we have $R\tilde R = -8(\partial_t\Omega/r)^3$, which is non-zero whenever there is a change in $\Omega$.

These give us a little more than zero intuition of what $R\tilde R$ designates. But how does the relation of this term to the chiral current shows itself in physical systems?

Quasi-particles propagating in a flowing fluid can be described as excitations of a field that lives on a curved background characterized by a non-trivial metric~\cite{Analogue}. One can intuitively guess that a rotating flow may translate, in analogue gravity language, to one with non-vanishing $R\tilde R$. On the other hand, in a rotating system of fermions an axial current emerges. This phenomenon is known by the name of Chiral Vortical Effect (CVE)~\cite{CVEAnatomy}. An interesting feature of this effect is its independent relation to temperature. For massless Dirac fermions, detached from chemical potentials (represented by $\mu$), this axial current relates to temperature $T$ by the following,
\begin{equation}
	\vec{j_5} = \vec{\Omega} \left( \frac{\mu}{4\pi^2} + \frac{|\vec{\Omega}|^2}{48\pi^2} + \frac{T^2}{12} \right) \, .
	\label{CVE}
\end{equation}
with $\Omega$ being the angular velocity of the rotating system. There are other similar effects to CVE that are worth mentioning such as the axial magnetic effect with measurable consequences where the same temperature dependence appears~\cite{AMEChernodub}. In all these cases the temperature dependence has been connected to the gravitational anomalies~\cite{GravAnomTrans} albeit they can be derived by other methods as well \cite{CVEStone}.

As the right hand side of \eqref{CVE}, apart from the geometric parameter $\vec{\Omega}$, contains also the temperature $T$, the winding density $R\tilde R$ must somehow encode the notion of temperature as well.
To see how these seemingly unrelated notions, namely gravity and temperature, can possibly be connected, its perhaps best to remember that gravity is the force that causes all types of \textit{energy} to flow; since it couples to everything. But this feature is just like temperature gradient. All energy carriers contribute to heat transfer. So there should be a relation, even though fictitious, between these two concepts, as has been employed before by Luttinger~\cite{LuttingerThermal} in his treatment of thermal transport using an auxiliary gravitational potential.

But how can we attribute a temperature to a specific space-time geometry? One way to have a notion of temperature associated to geometry is to look at black hole space-times. They radiate thermally. Let us for simplicity imagine a $(1+1)$ dimensional black hole system. An event horizon divides space-time into interior and exterior regions. In the exterior particles are doomed to move towards the \textit{future} and also are allowed to have positive or negative \textit{momenta}. In the interior the roles of space and time are ``swapped'', where now particles are doomed to go towards the \textit{singularity} and also are allowed to have positive or negative \textit{energy}. Having this permission (for negative energy particles in the interior) allows for a \text{real} pair creation near the horizon, where the negative energy particle is to be created inside the black hole while the positive exterior one can escape to infinity. In this way energy can be extracted from the black hole in the form of radiation. By calculating the probability of such pair creation \citep{Hawking,Parikh} we can see that this radiation is thermal with a temperature proportional to the surface gravity of the horizon---a completely geometrical quantity.

Therefore, for CVE, in order to get an anomalous contribution proportional to $\Omega$ and $T$, what we need is perhaps a geometry that has both a twist and a horizon. In fact we hope for a non-vanishing Pontryagin density from which a $\vec\Omega T^2$ term, as in equation \eqref{CVE}, can be extracted.

\subsection{Adding a Horizon}

The following metric \cite{CVEStone} has all the properties we are looking for, namely, a twist, a horizon, nonzero Pontryagin density, and asymptotic flatness as an additional welcoming feature. In return it is a bit more complicated than our previous examples:
\begin{align} \label{StoneMetric}
	ds^2 = &-f(z)\frac{\left(dt - \Omega r^2 d\phi \right)^2}{\left( 1 - \Omega^2 r^2 \right)} + \frac{1}{f(z)} dz^2 +dr^2 + \frac{r^2\left( d\phi - \Omega dt \right)^2}{\left( 1 - \Omega^2 r^2 \right)} \, . 
\end{align}
The horizon happens on $z=0$ where $f(z)$ is set to have a non-degenerate root. The Pontryagin density of the metric \eqref{StoneMetric} is given by
\begin{align}
	&\frac{1}{4}\epsilon^{\mu\nu\rho\sigma}R_{\mu\nu}^{\ \ \ \alpha\beta}R_{\rho\sigma\alpha\beta} = - \frac{ 2 \Omega f'(z) \left[ \left( 1-r^2 \Omega^2 \right)^2 f''(z)-8 \Omega^2 \left( 1-f(z) \right) \right] }{ (1- r^2 \Omega^2)^3 } \, . 
\end{align}
with the prime designating differentiation with respect to $z$. Recall that $\epsilon^{\mu\nu\rho\sigma}=\varepsilon^{\mu\nu\rho\sigma}/\sqrt{|g|}$ is the Levi-Civita \textit{tensor} and $|g|=r$ with our choice of metric.

One can effectively reduce the near horizon physics to that of a $(1+1)$ dimensional chiral quantum field theory~\citep{Wilczek}. The anomalous non-conservation relation for a current of either right or left moving particles is the same as equation \eqref{CVE} but now with half of the right hand side. Looking back at the anomalous relation with the above Pontryagin density, keeping only the leading term in $\Omega$ we have,
\begin{equation}
	\frac{1}{\sqrt{|g|}}\partial_z \left( \sqrt{|g|} j_5^z \right) = -\frac{\Omega}{192\pi^2}\partial_z \left(  f'(z)^2 \right) \, ,
\end{equation}
or
\begin{equation}
	j_5^z \bigg |_{z\rightarrow\infty} = \left. \frac{\Omega}{192\pi^2} f'(z)^2 \right|_{z=0}=  \frac{\Omega}{48\pi^2} \kappa^2 \, ,
\end{equation}
where $\kappa=f'(0)/2$ is the surface gravity of the horizon. Here we have used two boundary conditions: First, because of the asymptotic flatness, $f'(z)$ vanishes at infinity, second, a non-vanishing current at the horizon leads to infinite flux which we have abandoned on physical grounds~\cite{Wilczek2,Wilczek3}.

Temperature of the Hawking radiation out of the horizon is related to its surface gravity by $T_H=\kappa/2\pi$, in natural units. Through identifying this temperature with the temperature of our fermionic system we will have,
\begin{equation}
	j_5^z = \frac{1}{12}\Omega T_H^2 \, ,
\end{equation}
which is the temperature part of the CVE \eqref{CVE} as we were looking for.

\subsection{Turning Interactions On}

Looking back at \eqref{WardCurvedInt} we observe that had we introduced interactions at the beginning, we would have had a different coefficient behind the Pontryagin density,
\begin{align}
	\nabla_\mu j_5^\mu = \ \text{the electromagnetic term} + \frac{1}{384\pi^2 \left( 1+n_0 \lambda^2/\pi \right) }\epsilon^{\mu\nu\rho\sigma}R_{\mu\nu}^{\ \ \ \alpha\beta} R_{\rho\sigma\alpha\beta} \, ,
\end{align}
which would have been dragged all the way to the end result, without altering anything else. In that case we would instead get,
\begin{equation}
	j_5^z = \frac{1}{12 \left( 1+n_0 \lambda^2/\pi \right)} \Omega T_H^2 = \frac{1}{12}\Omega \tilde{T}_H^2 \, ,
\end{equation}

This can be interpreted in two ways: Either that the interactions alter the flow of charge along $z$, or, the temperature of this radiation is not simply given by $\kappa/2\pi$ in the presence of interactions. In the latter case the modified temperature would be $\tilde T_H \equiv T_H / \sqrt{1+ n_0 \lambda^2/\pi}$.

Mixed axial-gravitational anomalies similar to the non-conservation relation \eqref{WardCurved} have been invoked in the context of experimental condensed matter physics as in \cite{NatureGooth}. If these anomalies truly appear in thermal phenomena as they have in \cite{NatureGooth}, then according to what we have discussed so far, one expects them and their consequent thermal phenomena to be modified by the effects of interactions, for example, through equation \eqref{FullCurvedWard}. Seeing the modification in experiment would also be a confirmation on the previous results. On the other hand, if the modifications avoid observation then the association made between the gravitational anomaly and the thermal phenomena would go under question.

As the final remark here, let us go back to the black hole radiation. Energy and charge fluxes out of the black hole horizon have been evaluated before using gravitational anomalies~\citep{Wilczek2,Wilczek3}. What usually is employed as an underlying theory for calculating the Hawking radiation, is a non-interacting quantum field theory. It is therefore a notable question to ask how an interacting theory would differ in producing black hole radiation. Combining our method of treating interactions and the method used in~\citep{Wilczek2,Wilczek3} for calculating horizon fluxes in non-interacting theories, we can observe that depending on the type of interactions the resulting fluxes indeed can go under some modifications; even though perhaps interaction terms such as $\bar\psi\gamma^\mu\psi\bar\psi\gamma_\mu\psi$ are RG irrelevant.


\section{Beyond Electrical Interactions} \label{sec:Beyond}

The main concern of this paper has so far been the local current-current interactions of the general type of $\lambda^2_{\mu\nu} j^\mu j^\nu$. But not all interactions are between currents, or in other words, not all interactions are expressible in terms of electrical interactions. Take, for example, interactions between chiral currents $\lambda^2_{\mu\nu} j^\mu_5 j^\nu_5$. Even though in two spacetime dimensions this interaction is equivalent to the current-current interaction, as is discussed in the appendix \ref{sec:RemQues}, the equivalence does not hold in four dimensions. The spin-spin interactions between Dirac fermions can be seen as the spatial part of $j_\mu^5 j^\mu_5$. What follows is the investigation of chiral anomaly in presence of such an interaction.

The general local interaction between chiral currents,  $\lambda^2_{\mu\nu} j^\mu_5 j^\nu_5$, can be investigated through the same procedure established in previous sections, but for the sake of simplicity we are going to specialize to the Lorentz invariant case expressed by the following path-integral,
\begin{equation}
		I=\int \! \mathcal{D}[\bar{\psi}\psi] \exp i\left\{  \int \! d^4x \left[ \bar{\psi}i\slashed{D}\psi - \frac{1}{2} \lambda^2  j^\mu_5 j^5_\mu \right] \right\} .
		\label{InteractingIJ5}
\end{equation}
As before, it is possible to decouple the interaction term above by the help of an auxiliary field $s^\mu$,
\begin{equation}
		I \! = \! \int \! \mathcal{D}[\bar{\psi}\psi s_\mu] \exp  i\left\{\int \! d^4x \left[ \bar{\psi}i\slashed{D}_g\psi +\frac{1}{2}s_\mu s^\mu \right] \! \right\} ,
		\label{DecoupledIJ5}
\end{equation}
where now the generalized Dirac operator is given as $\slashed{D}_g \equiv \gamma^\mu\left( \partial_\mu -ieA_\mu - i\lambda s_\mu \gamma_5 \right)$. It is straightforward to check that a shift in the auxiliary field $s^\mu \rightarrow s^\mu - \lambda j^\mu_5$ decouples the auxiliary field from the fermions and leaves only a quadratic term in the action which can easily be integrated out to give \eqref{InteractingIJ5} back. But before we do that, we first benefit from the fact that the path-integral is formally diagonalized in its current state \eqref{DecoupledIJ5}, in which we can unambiguously calculate the chiral anomaly. (see subsection \ref{sec:Regularization} if needed.)

We encountered a constant axial-field appearing inside the generalized Dirac operator in previous sections. Being constant the axial-field would not contribute to the chiral anomaly. But here we are integrating over all configurations of $s_\mu$ in which case the constant $s_\mu$ becomes irrelevant. Therefore we need to recalculate the chiral anomaly in presence of both the gauge field $A_\mu (x)$ and the axial-field $s_\mu(x)$, which means we have to go over similar steps as in \eqref{2DCal}, when we traced over $\gamma_5$, but this time with the generalized Dirac operator that itself carries a $\gamma_5$ inside and also in four spacetime dimensions. This is a rather long and tedious task, so we skip to the result below that first appeared in~\cite{NonAbelianBosPRB} along with the corresponding Weyl anomaly.
\begin{align}
	  \partial_\mu j^\mu_5  =  \lambda\left[\frac{M^2}{2\pi^2 }+\frac{\lambda^2s_\mu s^\mu}{\pi^2}-\frac{\partial_\mu\partial^\mu}{12\pi^2}\right]\partial_\nu s^\nu 
	+\frac{\epsilon^{\mu\nu\rho\sigma}}{16\pi^2}\left[\frac{\lambda^2}{3} G_{\mu\nu}G_{\rho\sigma}
+e^2 F_{\mu\nu}F_{\rho\sigma}\right] \, ,
    \label{gamma5}
\end{align}
with $G_{\mu\nu} \equiv \partial_\mu s_\nu - \partial_\nu s_\mu$ and $M$ defined in \eqref{RegularizationIntro}. The terms constructed out of $s_\mu$ in the above equation are all odd in each component of $s_\mu$. Therefore, if we shift $s^\mu$ by its on-shell value, $-\lambda j^\mu_5$, and consequently make the action quadratic in $s_\mu$, all the odd terms in above will vanish after the integration. Therefore we will have for the chiral anomaly,
\begin{align}
	\partial_\mu j^\mu_5 =-\lambda^2\left[\frac{\tilde{M}^2}{2\pi^2}+\frac{\lambda^4j^5_\mu j_5^\mu}{\pi^2}-\frac{\partial_\mu\partial^\mu}{12\pi^2}\right]\partial_\nu j_5^\nu 
	+\frac{\epsilon^{\mu\nu\rho\sigma}}{16\pi^2}\left[\frac{4\lambda^4}{3} \partial_\mu j^5_\nu \partial_\rho j^5_\sigma
    +e^2 F_{\mu\nu}F_{\rho\sigma}\right] \, ,
\label{AnomalyJ5}
\end{align}
with $\tilde{M}$ now containing also the constant contribution coming from $\langle s_\mu s^\mu \rangle$ in addition to $M$. Depending on the physics behind the anomalous relation above, we adjust the chiral symmetry breaking formula. For example, if $j^\mu_5$ is really describing the spin on a Dirac fermions, then $j^5_\mu j_5^\mu$ becomes a constant $S$ which renders \eqref{AnomalyJ5} as
\begin{align}
	\partial_\mu j^\mu_5 =\left[1+ \lambda^2\left(\frac{ \tilde{M}^2}{2\pi^2}+\frac{\lambda^4 S}{\pi^2}-\frac{ \Box}{12\pi^2}\right)\right]^{-1} \times \frac{\epsilon^{\mu\nu\rho\sigma}}{16\pi^2}\left[\frac{4\lambda^4}{3} \partial_\mu j^5_\nu \partial_\rho j^5_\sigma
+e^2 F_{\mu\nu}F_{\rho\sigma}\right] \, .
 \label{AnomalySpin}
\end{align}

\section{Conclusion}

Interactions, as we saw, have non-trivial effects on the chiral anomaly and its many consequences. The non-conservation of the chiral current is modified by the presence of interactions and consequently also all the phenomena associated to it: The Hall conductivity in the inhomogeneous and non-equilibrium limit; the electric charge of $(1+1)$-dimensional pseudo-particles; the existence of longitudinal non-equilibrium Hall conductivity; the density response to magnetic field; the chiral magnetic effect; the chiral vortical effect and thermal or geometrical responses are the few examples we discussed here explicitly. We also found that the interplay of anomaly with interactions and Weyl separation leads to the existence of curious massive modes coupled to an axion-electromagnetic field. Lastly we showed that the appearance of these modifications are not subject to only electrical interactions but can easily occur under the influence of other types such as spin-spin interactions. In order to make certain that the regularization used for the calculation of the anomaly is indeed justified we have developed a general path-integral regularization procedure that reduces to well-known regularizations such as that of Fujikawa's or Pauli-Villars in those cases that these regularizations become relevant.\footnote{The details of this construction is presented in appendix~\ref{sec:SelfRegularization}.}

The exactness of chiral anomaly extends to interacting systems as well, providing rare non-perturbative results. One can hope that the results and the method set forth in this paper will extend to various other problems as well. Possible examples include the Weyl-Kondo semimetal~\cite{NonAbelianBosPRB}, interacting topological insulators~\cite{TopolInsInt}, studies on Hawking radiation via anomalies~\cite{Wilczek} in interacting field theories, non-equilibrium studies of Hall phenomena, and can even extend to odd spacetime dimensions~\cite{TopolCriterion}. For our investigations here we used the path-integral technique for the low energy effective theory. It is nonetheless desirable to understand how these effects arise from a microscopic lattice model and to observe how they coincide in the low energy limit as the extra symmetries of the effective theory emerge.

\acknowledgments
This work was supported by the U.S. Department of Energy, Office of Science, Basic Energy Sciences under Award No. DE-SC0001911 and Simons Foundation (V.G) and through ERC under Consolidator grant number 771536 NEMO (CR).


\bibliographystyle{JHEP}
\bibliography{mybib.bib}


\appendix

\section{A Calculation Without Auxiliary Fields}

Calculating the effects of interactions on anomalous relation through an auxiliary field is insightful on its own, but as additional clarification we present a slightly different calculation here, which does not rely on auxiliary fields. We begin by dividing the spinor fields into left and right moving parts,
\begin{equation}
\begin{split}
	&\psi_R = \frac{1+\gamma_5}{2} \psi,   \ \ \ \   \bar{\psi}_R = \bar{\psi} \frac{1 - \gamma_5}{2}, \\
	&\psi_L = \frac{1-\gamma_5}{2} \psi,   \ \ \ \   \bar{\psi}_L = \bar{\psi} \frac{1 + \gamma_5}{2}.
\end{split}
\end{equation} 
This way both the electrical current and the chiral current will also be divided into $L$ and $R$ parts. Therefore, the current-current interaction (which is nothing but a density-density interaction made Lorentz invariant) is given by
\begin{equation}
	-\frac{\lambda^2}{2} \bar{\psi}\gamma^\mu\psi\bar{\psi}\gamma_\mu\psi = -\frac{\lambda^2}{2} \left( j_R^\mu j^R_\mu  +  j_L^\mu j^L_\mu + 2 j_R^\mu j^L_\mu \right) \, ,
\end{equation}
where $j^\mu_{R,L} \equiv \bar{\psi}_{R,L}\gamma^\mu\psi_{R,L}$ and we used the fact that paired Grassmann numbers are commutative. The quantum system is therefore described by the following path-integral,
\begin{equation}
	I = \int \mathcal{D}\bar{\psi}_R \mathcal{D}\psi_R \mathcal{D}\bar{\psi}_L \mathcal{D}\psi_L e^{iS[\bar{\psi}_R,\psi_R,\bar{\psi}_L,\psi_L,A_\mu]} \, .
\end{equation}
The action and the path-integral are both symmetric under $ R \leftrightarrow L$ transformation and the action $S$ for the (1+1)-dimensional case is given by
\begin{align}
	S = \int d^2x \bigg [ &\bar{\psi}_R i \gamma^\mu D_\mu \psi_R + \bar{\psi}_L i \gamma^\mu D_\mu \psi_L -\frac{\lambda^2}{2} \left( j_R^\mu j^R_\mu  +  j_L^\mu j^L_\mu + 2 j_R^\mu j^L_\mu \right) \bigg ] \, ,
\end{align}
with $D_\mu \equiv \partial_\mu - ieA_\mu $.

By an infinitesimal chiral rotation in either right or left moving pairs we can obtain corresponding anomalous relations. Since we have divided our original fields into two distinct $R$ and $L$ parts, the anomalous term, which was originally equal to the difference of right and left moving zero modes $n_+ - n_-$, gets divided by a factor of two; so that in the absence of interactions the anomalous relation becomes
\begin{equation}
	\partial_\mu {j_5^\mu}_{R,L} = \frac{e}{2\pi}\epsilon^{\mu\nu}\partial_\mu A_\nu 
\end{equation}
instead of
\begin{equation}
	\partial_\mu {j_5^\mu} = \partial_\mu {j_5^\mu}_R + \partial_\mu {j_5^\mu}_L  = \frac{e}{\pi}\epsilon^{\mu\nu}\partial_\mu A_\nu \, .
	\label{RplusL}
\end{equation}
Since $\psi_R$ pairs are independent from $\psi_L$ pairs, each time we chirally rotate one pair we can treat the other as constant fields and its corresponding current a constant four-vector. For example, let us rotate $\psi_R$ to $e^{i\alpha\gamma_5} \psi_R$ while we have prepared the action in the following form,
\begin{equation}
\begin{split}
	S = \int \! d^2x \bigg [ \bar{\psi}_R i \gamma^\mu \left( D_\mu + i \lambda^2 j^L_\mu \right) \psi_R &+ \mathcal{L} [\psi_L,\bar{\psi}_L] \bigg ] \, .
\end{split}
\end{equation}
We see that $j^L_\mu$ is now acting as an external field for right moving fermions. (And it does so through its field strength $F^{j_L}_{\mu\nu} \equiv \partial_\mu j^L_\nu - \partial_\nu j^L_\mu$.) 

Now let us assume that in the presence of interactions the total anomalous term, i.e. the right hand side of equation \eqref{RplusL}, will be modified by a constant factor $\kappa(\lambda^2)$ where $\lambda^2/2$ is the interaction coupling:
\begin{equation}
	\partial_\mu {j_5^\mu} = \kappa(\lambda^2) \frac{e}{\pi}\epsilon^{\mu\nu}\partial_\mu A_\nu \, .
	\label{TotalWard}
\end{equation}
For the right and left moving anomalous relations the interaction term is divided into these two parts as well. Note that before, the whole field, containing both the right and left moving modes, was interacting with itself. But now only half of that self-interaction appears as a self-interaction for each left and right moving field, while the other half appears as the interaction between one field and the other. Therefore, the corresponding coefficient for their anomalous term will be $\kappa(\lambda^2/2)$ instead of $\kappa(\lambda^2)$. All in all, the anomalous identity would then be
\begin{equation}
	\partial_\mu {j_5^\mu}_R = \kappa\left(\frac{\lambda^2}{2}\right) \left [ \frac{e}{2\pi}\epsilon^{\mu\nu}\partial_\mu A_\nu - \frac{\lambda^2}{2\pi} \epsilon^{\mu\nu}\partial_\mu j^L_\nu \right ] \, .
\end{equation}

Since the transformation $R \leftrightarrow L$ is a symmetry of the path-integral, we can write down the same relation for left moving part of the chiral current. One can then sum up these relations to obtain the corresponding anomalous identity for the total chiral current: 
\begin{equation}
	\partial_\mu j^\mu_5 = \kappa\left(\frac{\lambda^2}{2}\right) \left [ \frac{e}{\pi}\epsilon^{\mu\nu}\partial_\mu A_\nu - \frac{\lambda^2}{2\pi} \partial_\mu j^\mu_5 \right ] \, ,
\end{equation}
where we have used the relation $\epsilon^{\mu\nu}\gamma_\nu = \gamma^\mu\gamma_5$ to change $j^\mu$ to $j_5^\mu$. Algebraic rearrangements then give,
\begin{equation}
	\partial_\mu j^\mu_5 = \frac{\kappa(\lambda^2/2)}{1+\frac{\lambda^2}{2\pi}\kappa(\lambda^2/2)} \frac{e}{\pi}\epsilon^{\mu\nu}\partial_\mu A_\nu \, .
\end{equation}

Looking at equation \eqref{TotalWard} prompts a consistency equation:
\begin{equation}
	\kappa (\lambda^2) = \frac{\kappa(\lambda^2/2)}{1+\frac{\lambda^2}{2\pi}\kappa(\lambda^2/2)} \, .
\end{equation}
In terms of $\kappa(\lambda^2)^{-1}$ the form of the above equation becomes simpler,
\begin{equation}
	\kappa\left(\lambda^2\right)^{-1} = \kappa \left( \lambda^2/2 \right)^{-1} + \frac{\lambda^2}{2\pi} \, .
\end{equation}
Knowing that $\kappa(0)$ is equal to one, this equation has a unique solution: $\kappa(\lambda^2)^{-1}= 1 +\lambda^2/\pi$. By this we eventually arrive at the final stage of equation \eqref{TotalWard},
\begin{equation}
	\partial_\mu {j_5^\mu} = \frac{1}{1+\lambda^2/\pi} \frac{e}{\pi}\epsilon^{\mu\nu}\partial_\mu A_\nu \, .
\end{equation}


\section{1+1 Dimensions Remaining Remarks} \label{sec:RemQues}

Here we answer two simple questions that are prone to be asked regarding the decoupling procedure employed in our calculation, and another possible four fermionic interaction.

There are different ways we can decouple the current-current term via a Hubbard-Stratonovich field $a_\mu$. One might then tend to ask whether these different ways will lead to different answers or not; if they do the result would have been unphysical, but this is not the case. We can verify this through the simple example below which is generalizable to more complicated cases that appears in later sections when we discuss the four dimensional case. Consider the action,
\begin{equation}
	S=\int \! d^2x \left[ \bar{\psi}i\gamma^\mu (\partial_\mu -ieA_\mu -i \text{L} a_\mu ) \psi + \frac{\Gamma}{2}a_\mu a^\mu \right].
\end{equation}
On-shell value of $a^\mu$ is $-(\text{L}/\Gamma) \bar{\psi}\gamma^\mu\psi$. After a shift in the auxiliary field by its on-shell value and integrating the shifted auxiliary field out, the action turns out as below,
\begin{equation}
	S=\int \! d^2x \left[ \bar{\psi}i\gamma^\mu (\partial_\mu -ieA_\mu  ) \psi  -\frac{\text{L}^2}{2\Gamma} \bar{\psi}\gamma^\mu\psi\bar{\psi}\gamma_\mu\psi \right].
\end{equation}
We set $\text{L}^2/\Gamma \equiv \lambda^2$ to make the action look like \eqref{2DInetractingAction}. On the other hand, the anomalous identity reads,
\begin{equation}
	\partial_\mu j^\mu_5 = \frac{e}{\pi}\epsilon^{\mu\nu}\partial_\mu A_\nu + \frac{\text{L}}{\pi}\epsilon^{\mu\nu}\partial_\mu a_\nu \, ,
\end{equation}
which after the shift and integration of the auxiliary field becomes,
\begin{equation}
	\partial_\mu j^\mu_5 = \frac{e}{\pi}\epsilon^{\mu\nu}\partial_\mu A_\nu - \frac{\text{L}^2}{\Gamma\pi}\epsilon^{\mu\nu}\partial_\mu a_\nu ,
\end{equation}
which gives back
\begin{equation}
	\partial_\mu j^\mu_5 =\frac{1}{1+\lambda^2/\pi}\frac{e}{\pi}\epsilon^{\mu\nu}\partial_\mu A_\nu \, ,
\end{equation}
regardless of how the decoupling of the interaction term is conducted.

Another question is regarding other possible interaction terms. One candidate is $j^\mu_5 j_\mu^5$; but in two spacetime dimensions this interaction is no different than $j^\mu j_\mu$. One can also consider interaction terms that do not respect the chiral symmetry such as $(\bar{\psi}\psi)^2$. The fact that it manifestly breaks the chiral symmetry does not readily mean that it will not alter the non-trivial Jacobian of the transformation.

The path-integral under consideration would be
\begin{equation}
	I = \int  \mathcal{D}\bar{\psi} \mathcal{D}\psi  \exp\int d^2x \left [ \bar{\psi} i \gamma^\mu D_\mu \psi + \Lambda^2 (\bar{\psi}\psi)^2 \right ] ,
	\label{psi4}
\end{equation}
which by using an auxiliary field $\phi$ can also be written in the following form,
\begin{align}
		& \int \mathcal{D}\bar{\psi} \mathcal{D}\psi \mathcal{D}\phi \exp \int \! d^2x \left [ \bar{\psi} i \gamma^\mu D_\mu \psi + 2\Lambda \phi \bar{\psi}\psi - \phi^2 \right ] \nonumber \\
		&= \int \mathcal{D}\bar{\psi} \mathcal{D}\psi \mathcal{D}\phi \exp \int \! d^2x \left [ \bar{\psi} i (\slashed{D} - 2i\Lambda\phi) \psi  - \phi^2 \right ] .
\end{align}
Either by looking at the first line in the above and see that all we have done is introducing a mass term, or by looking at the second line and following strict calculations of anomaly, we can see that the Jacobian of chiral transformation $\psi \rightarrow e^{i\alpha\gamma_5}\psi$ is not modified by this term. Nevertheless, being a mass term, it does not remain invariant under a chiral transformation, and contributes to the non-conservation as below:
\begin{equation}
	\langle \partial_\mu j_5^\mu \rangle_\phi = \mathcal{A}_5 + \langle 4i\Lambda\phi\bar{\psi}\gamma_5\psi \rangle_\phi \, ,
\end{equation}
where $\mathcal{A}_5$ is the usual anomalous term coming from the Jacobian and there is a remaining integration over the $\phi$ field. The end result after the integration is
\begin{equation}
	\partial_\mu j_5^\mu = \mathcal{A}_5 +  4i\Lambda^2\bar{\psi}\psi \bar{\psi}\gamma_5\psi \, ,
\end{equation}
which agrees with a Ward identity derived directly from equation \eqref{psi4}.


\section{Path-Integral Self Regularization}
\label{sec:SelfRegularization}

Unlike regular path-integration, fermionic path-integrals are defined by left differentiation of Grassmann numbers and they need regularization. For example, path integration over fermionic degrees of freedom of a free massive Dirac fermion theory yields the determinant of the Dirac operator $\det (i\slashed{D} + m) = \prod l_n$ only if the product is well-defined which is the case when $\sum_n^\infty |l_n -1|$ converges~\cite{InfiniteProduct}. But eigenvalues, $l_n$, of the Dirac operator are not even bounded. Thus the well-defined path-integral must carry a type of regularization with itself. This is where action functional intrudes into measure's business. The path-integral (or the partition function) has two elements; the measure and the action; the regularization is not present in the latter. But \textit{how} the measure is regularized must only be determined by the action, since the partition function is to be a self-sufficient object and should be indifferent about the backstory of why it is written. Consequently, the regularization, being determined only by the action, will share the symmetries of the action; even if the measure does not respect them all.

For the sake of clarification let us investigate an instance of a regularization which solely relies on the path-integral itself. Similar to \eqref{SpinorExpansion} we first expand the fermionic fields in the basis of a generic hermitian operator. Then the action will be a function of Grassmann numbers $\bar{b}_n$ and $a_n$.
\begin{equation}
	\int \mathcal{D}\bar{\psi}\mathcal{D}\psi e^{S[\bar{\psi},\psi]} \rightarrow \int \prod_n d\bar{b}_n da_n \mathcal{F}\left[ \{ \bar{b}_n\} , \{a_n\} \right] \, ,
\end{equation}
which is not yet regularized. $\mathcal{F}$ is generally written as,
\begin{equation}
	\mathcal{F} =\sum_{k=0}^\infty \sum_{i_1, \cdots ,i_{2k}} c^{i_1 \cdots i_k i_{k+1} \cdots i_{2k}} \bar{b}_{i_1} \cdots \bar{b}_{i_k} a_{i_{k+1}} \cdots a_{i_{2k}} \, ,
	\label{Grassmann}
\end{equation}
with $c$ being complex completely anti-symmetric tensors. We can then define a complex number $\ell_m$ as follows,
\begin{equation}
	\ell_m \equiv \frac{\int \prod d\bar{b}_n da_n \mathcal{F}\left[ \{ \bar{b}_n\} , \{a_n\} \right]}{\int \prod d\bar{b}_n da_n \bar{b}_m a_m \mathcal{F}\left[ \{ \bar{b}_n\} , \{a_n\} \right]} \, ,
	\label{EigenIntegral}
\end{equation}
given directly by the path-integral and not some specific building block of it. The reason for this definition becomes clearer when when we look at a specific class of $\mathcal{F}$ given as
\begin{equation}
    \mathcal{F}_D = \prod_n \left( 1 + l_n \bar{b}_n a_n \right) = \exp\left\{\sum_n l_n \bar{b}_n a_n\right\} \, .
    \label{DiagonalF}
\end{equation}
In this situation we say $\mathcal{F}_D$ is formally diagonalized in terms of Grassmann numbers $\{\bar{b}_n\}$ and $\{ a_n \}$. Then $\ell_m$ defined as \eqref{EigenIntegral} is equal to $l_m$ from the above. Moreover, a formally diagonalized $\mathcal{F}_D$ allows the path-integral to be completely factorized as $\int \prod d\bar{b}_n d a_n \exp\{l_n \bar{b}_n a_n \}$. Having the path-integral written in this way makes it easy to set a cut-off in $l_n$ for the modes of integration, leading to a natural regularization. For general $\mathcal{F}$ which is not formally diagonalized $\ell_n$ plays the role of $l_n$ and can be used for the regularization process instead.

We can then form the product
$$\mathcal{I} = \prod_n f_M(\ell_n) \, ,$$
as a regularized value for the path-integral, with $f_M(x)$ being a function that rapidly declines to unity for $x^2>M^2$ and $f_M(x)=x$ for $x^2<M^2$.

In other words, we can simply truncate the path-integration by limiting it to only those modes whose corresponding $\ell^2_n$ is less than a certain value $M^2$, which can also be done smoothly by weighting each mode of integration $\int d\bar{b}_n da_n$ by a coefficient $f_n$ which is equal to one for $\ell^2_n < M^2$, and smoothly but rapidly approaches to $\ell_n^{-1}$ for $\ell^2_n > M^2$. We thus obtain a well-defined fermionic path-integral.

One might then wonder if this approach include any other already known regularizations; the answer is yes. This of course includes the regularization used in Fujikawa's approach, as we will see later, but is more general than that. To see this, consider the case where we have chosen $f_M(x)$ to be $(1+iM/x)^{-1}$:
\begin{align}
	f_M(\ell_m)^{-1} & =\frac{\ell_m + iM}{\ell_m} = 1+ \frac{iM}{\ell_m}  \nonumber \\
	&= \frac{\int \prod d\bar{b}_n da_n \mathcal{F}_\mathbf{n}}{\int \prod d\bar{b}_n da_n \mathcal{F}_\mathbf{n}} + \frac{\int \prod d\bar{b}_n da_n \left( iM \bar{b}_m a_m \right) \mathcal{F}_\mathbf{n}}{\int \prod d\bar{b}_n da_n \mathcal{F}_\mathbf{n}}  \nonumber \\
	&=\frac{\int \prod d\bar{b}_n da_n \exp\left(iM\bar{b}_m a_m \right) \mathcal{F}_\mathbf{n}}{\int \prod d\bar{b}_n da_n \mathcal{F}_\mathbf{n}} \, ,
\end{align}
where now the subscript $\mathbf{n}$ in $\mathcal{F}_\mathbf{n}$ designates the dependence of $\mathcal{F}$ on $\{\bar{b}_n\}$ and $\{a_n\}$, shortening the previous notation. Therefore, we can define the product $\mathcal{I}$ as
\begin{align}
	 \prod_m \frac{\ell_m}{\ell_m + iM} & = \frac{\int \left[ \prod_m \left( \prod_n d\bar{b}_n^m da_n^m \right) \mathcal{F}_\mathbf{n}^m \right] }{\int \left[ \prod_m \left( \prod_n  d\bar{b}_n^m da_n^m \right) e^{iM\bar{b}_m^{\tilde{m}}  a_m^{\tilde{m}} } \mathcal{F}_\mathbf{n}^m \right]} \nonumber \\
	& = \!  \frac{\int \left[ \prod_m \left( \prod_n d\bar{b}_n^m da_n^m \right) \mathcal{F}_\mathbf{n}^m \right] }{\int \left[ \prod_m \left( \prod_n  d\bar{b}_n^m da_n^m \right)  \mathcal{F}_\mathbf{n}^m \right] \exp \{iM\sum _m \bar{b}_m^{\tilde{m}}  a_m^{\tilde{m}} \} } .
\end{align}
Here each $( \{ \bar{b}_n \}^m , \{ a_n \}^m)$ is a distinct copy of $( \{ \bar{b}_n \}, \{ a_n \})$ for distinct values of $m$, each of which span a similar space of Grassmann numbers with the same dimension. The number of copies (domain of $m$) is equal to the dimensionality of each copy (domain of $n$). Also, $\tilde{m}(m)$ is any permutation of $m$.

The above yields a regularized path-integral without any external information which should be regarded irrelevant to the path-integral as a mathematical object. The term inside the exponential in the last line, is a sum over all possible values of $m$, hence $m$ is a dummy index and can be substituted with any other letter. On the other hand, the form of $\mathcal{F}^m_\mathbf{n}$ is the same for all $m$. Therefore, considering the Grassmann algebra, one can separate the subspace spanned by $\left( \{ \bar{b}^{\tilde{m}}_m \} , \{a^{\tilde{m}}_m \} \right)$ from the rest of Grassmann space, taking a quotient of $\prod_m \mathcal{F}^m_\mathbf{n}$ without changing the ultimate value of the product, and rewrite it as,
\begin{align}
	\mathcal{I}  =  \frac{\mathcal{Q} \times \int \left( \prod_n d\bar{b}^{\tilde{n}}_n da^{\tilde{n}}_n \right) \mathcal{F}_{\mathbf{n}}^{\tilde{n}} }{\mathcal{Q} \times \int \left( \prod_n  d\bar{b}_n^{\tilde{n}} da_n^{\tilde{n}} \right)  \mathcal{F}_\mathbf{n}^{\tilde{n}} \exp \{iM\sum _n \bar{b}_n^{\tilde{n}}  a_n^{\tilde{n}} \} } \, ,
\end{align}
where the quotient $\mathcal{Q}$ is what remains from both numerator and denominator when other terms are extracted out, and $\mathcal{F}^{\tilde{n}}_{\mathbf{n}} \equiv \mathcal{F}\left[ \{\bar{b}^{\tilde{n}}_n \}, \{ a^{\tilde{n}}_n \} \right]$. After canceling $\mathcal{Q}$ the nominator will be the original path-integral we began with while the denominator is the same path-integral with a mass term $i\int d^2x M\bar{\psi}\psi$ introduced in its action. Therefore, the product $\mathcal{I}$ can also be written as one path-integral,
\begin{equation}
	\mathcal{I} = \int \mathcal{D}\bar{\psi}\psi\mathcal{D}\bar{\varphi}\varphi
	\mbox{\large{ $e^{S[\bar{\psi},\psi] + S[\bar{\varphi},\varphi] + i\int d^2 x M \bar{\varphi}\varphi }$ }} \, ,
\end{equation}
which is the famous Pauli-Villars regularization. Notice that $S[\bar{\varphi},\varphi]$ has the similar form as $S[\bar{\psi},\psi]$ but with spinor fields substituted by bosonic fields $\bar{\varphi}(x)$ and $\varphi(x)$. Also recall that Grassmann integral is defined by left derivative and therefore appears inversely to what a bosonic integral would.

Thus the regularized fermionic path-integral introduced here is a general case that includes, but is not limited to, the Fujikawa and Pauli-Villars regularizations.

As an interesting remark, let us now look at a global chiral rotation of both fermionic \eqref{ChiralTransG} and bosonic fields,
\begin{align}
	\bar{\varphi} \rightarrow \bar{\varphi}e^{i\alpha\gamma_5}, \ \ \varphi \rightarrow e^{i\alpha\gamma_5} \varphi \, ,
\end{align}
for an angle $\alpha = \frac{\pi}{2}$. Jacobian of transformation for fermionic fields is inversely equal to that of bosonic fields, therefore, the whole measure $\mathcal{D}\bar{\psi}\psi\mathcal{D}\bar{\varphi}\varphi$ remains preserved. But the same is not true for the action. The effect of this transformation on the action is that the mass term obtains a minus sign. Thus,
\begin{align}
	- \delta \ln \mathcal{I} &= \sum_n \ln \frac{\ell_n}{\ell_n + iM} - \sum_n \ln \frac{\ell_n}{\ell_n - iM} = 2i \operatorname{Im} \mathcal{I} \nonumber \\
	& = \sum_n \ln \frac{\ell_n - iM}{\ell_n + iM} \approx - i\pi \left( \sum_{\ell_n>0} 1 - \sum_{\ell_n <0} 1 \right) \, ,
\end{align}
for the large $M$ limit. But if there were no bosonic fields, the Jacobian of fermionic measure \eqref{lnJacobian}, would have contributed in the exact same way according to \eqref{Index}. Hence, $-\frac{2}{\pi} \operatorname{Im}\mathcal{I}$ is the corresponding path-integral version of the index \eqref{Index}. Notice that it is $\ell_n$, not eigenvalues of an externally given operator, that appears in the index theorem, and thus in the anomalous term as well.

We saw that our regularization indeed comes down to well-known forms at special cases, however, $\ell_n$, when isolated, is only well-defined for certain bases, which our generic basis is assumed to be one of them. From a different perspective, therefore, the question of finding a basis that makes definition \eqref{EigenIntegral} realizable, leads to the set of path-integral elected bases. We could have started by this question instead, namely, what basis is preferred by the path-integral. As an answer we can ask for a basis $\{ \bar{b}_n , a_n\}$ that turns the unregularized fermionic path-integral into products of separate Grassmann integrations,
\begin{equation}
	\int \mathcal{D}\bar{\psi}\mathcal{D}\psi e^{S[\bar{\psi},\psi]} \rightarrow \prod_n \left( \int d\bar{b}_n da_n F_n \left[ \bar{b}_n, a_n \right] \right) \, ,
	\label{ProdPathIntegral}
\end{equation}
with $\mathcal{F} = \mathcal{F}_D \equiv \prod_n F_n$ so that \eqref{ProdPathIntegral} is a mere rearrangement. Since each term after integration will be a complex number, there exists a natural regularization:
\begin{equation}
	\int \mathcal{D}\bar{\psi}\mathcal{D}\psi e^{S[\bar{\psi},\psi]} \equiv \prod_n f_M\left( \int d\bar{b}_n da_n F_n \left[ \bar{b}_n, a_n \right] \right) \, ,
	\label{RegularizationBasis}
\end{equation}
with $f_M$ defined as before. Again, this leads to a well-defined path-integral.

Finding such a basis is subject to solving the following ``eigenvalue'' equation,
\begin{equation}
	  \lim_{K \to \infty} \left( \int d\bar{b}_n da_n  \left[ \mathcal{F}_{\Lambda_K} \right] - \ell_n \mathcal{F}_{\Lambda_{K-1}} \right) = 0 \, ,
\end{equation}
where $\mathcal{F}_{\Lambda_K}$ is defined as \eqref{Grassmann} when variable $k$ spans from $0$ to $K$. $\Lambda_K$ is the space spanned by $k$ pairs of Grassmann variables, and $\Lambda_{K-1}$ is the subspace of $\Lambda_K$ with $\bar{b}_n$ and $a_n$ excluded. At certain times, the above equation reduces to
\begin{equation}
	(c_n + \bar{b}_n a_n)  \int d\bar{b}_n da_n   \left[ \mathcal{F} \right] = \ell_n \mathcal{F} \, .
\end{equation}
It is then easy to see that we can equivalently write down the following path-integral
\begin{equation}
	\int \mathcal{D}\bar{\psi}\mathcal{D}\psi e^S \rightarrow \int \prod_n d\bar{b}_n da_n \exp\left\{\sum_n \ell_n \bar{b}_n a_n\right\} \, ,
	\label{FormallyDiagonalized}
\end{equation}
where we can say that the action is formally diagonalized in the chosen basis.

As a conclusion we can say that action determines what bases are preferable for the path-integral to be regularized with respect to. This means that the measure, which is regularized in the same manner, is affected by the action, or the weight, of the fermionic path-integral. Therefore, adding interaction terms to the action does not necessarily leave the measure, and the anomaly that comes from it, untouched. At last, we saw that a way to correctly see how the measure is affected is to move to a basis that formally diagonalizes the action.


\end{document}